%% file: HK-triangulation.tex
\documentclass[12pt]{article}

\usepackage{psfig}
\usepackage{epsfig, graphics}
\usepackage{amsmath, amstext,amssymb}
\usepackage{algorithm}
\usepackage{algorithmic}

\usepackage{enumerate, xspace}

\newcommand{\cali}[1]{{\cal #1}}
\newcommand{\ie}{{i.e.}}
\newcommand{\ab}{\allowbreak }
\newcommand{\vect}[1]{{\bf #1}}
\newcommand{\R}{\mathbb{R}}

\newcommand{\Rn}[1]{\mathbb{R}^{#1}}
\newcommand{\RnR}[1]{(\mathbb{R}^{#1}\times\mathbb{R})}

\newcommand{\ignore}[1]{}%

\newcommand{\trc}[1]{T_{#1}}
\newcommand{\ndim}[1]{${#1}$-{\rm dimensional}\xspace}

\newtheorem{corollary}{Corollary}

 \newtheorem{definition}{Definition}
\newtheorem{example}{Example}

\newtheorem{property}{Property}

\newcommand{\st}{{\rm spatio-temporal}\xspace}
\newcommand{\sti}{\emph{spatio-temporal}\xspace}

\def\squareforqed{\hbox{\rlap{$\sqcap$}$\sqcup$}}
\def\qed{\ifmmode\squareforqed\else{\unskip\nobreak\hfil
\penalty50\hskip1em\null\nobreak\hfil\squareforqed
\parfillskip=0pt\finalhyphendemerits=0\endgraf}\fi}

\begin{document}

\title{An Affine-invariant \\
Time-dependent Triangulation \\ of
Spa\-tio-tem\-po\-ral Data\footnote{An extended abstract of an earlier version of this paper appeared as~\cite{kh-gis04}.}}

\author{
Sofie Haesevoets and Bart Kuijpers\footnote{Contact author: {\tt bart.kuijpers@uhasselt.be}}\\
Hasselt University, Belgium}
\date{}
\maketitle

\begin{abstract}
In the \emph{geometric data model} for spatio-temporal data, introduced by Chomicki and Revesz~\cite{cr-99},
spa\-tio-tem\-po\-ral data are modelled as
a finite collection of triangles that are transformed
by time-dependent affinities of the plane.
To facilitate querying and animation of spa\-tio-tem\-po\-ral data,
we present a \emph{normal form} for data in the geometric data model.
We propose an algorithm for constructing this normal form
via a \emph{spa\-tio-tem\-po\-ral
triangulation} of geometric data objects.
This triangulation algorithm generates new geometric data objects
that partition the given objects both in
space and in time.
A particular property of the proposed partition is that it is
\emph{invariant under time-de\-pendent affine
transformations}, and hence independent of
the particular choice of  coordinate system used to describe 
the spatio-temporal data in. We
can show that our algorithm works correctly and has a polynomial  time
complexity (of reasonably low degree in the number of
input triangles and the maximal degree of the polynomial functions that describe the transformation functions).
We also discuss several
possible applications of this spa\-tio-tem\-po\-ral triangulation.
\end{abstract}




\section{Introduction and Summary}\label{intro}
At this moment, spatial databases are a well-established area of
research. Since most natural and man-made phenomena have a
temporal as well as a spatial extent a
lot of attention has also been paid, during the last decade, to modelling and querying
spa\-tio-tem\-po\-ral
data~\cite{STDBM99,cz-00,amai03,CHOROCHRONOS,grumbach,ge-00,kpv-00,req}.
Several data models for representing spa\-tio-tem\-po\-ral data
have been proposed already. In this article, we adopt the
\emph{geometric data model} that was introduced by Chomicki and
Revesz~\cite{cr-99} and of which closure properties under boolean operations were 
later studied by them and the present
authors~\cite{amai03,hk-00}. In the geometric data model,
\emph{spa\-tio-tem\-po\-ral objects} are finitely represented as
\emph{geometric objects}, which in turn are collections of atomic
geometric objects. An atomic geometric object is given by its
spatial reference object (which determines its shape), a time
interval (which specifies its lifespan) and a transformation
function (which  determines the movement of the object during the
time interval).

Although this model is very natural, it is not  immediately clear how the \st object represented by
some atomic geometric objects looks like. This difficulty has several possible reasons.
To start with, the time domain of the \st object has to be computed from the time domains of
all atomic geometric objects, which might overlap or may contain gaps (\ie, moments when the \st object
does not exist). Furthermore, two different sets of atomic geometric objects may represent the same \st object and there may
be elements in the set of atomic geometric objects that do not contribute to the \st object at all, as they
may be overlapped totally by other atomic objects. In short, the proposed geometric data model would benefit from a
\emph{normal form} that supports visualization and describes the objects in a unique way.

We propose as a normal form an \emph{affine-invariant spa\-tio-tem\-po\-ral triangulation}. This triangulation can be used to pre\-pro\-cess geometric objects
in order to facilitate querying and animation in such a way
that queries can be executed much more efficiently and require few additional computations.
The main reason for this is that in a spatio-temporal triangulation  the data is partitioned in space as well as in time. Hence, no objects overlap, thus
reducing unnecessary computations.
Actually, we deviate a little from the strict mathematical concept of a triangulation and allow triangles in a  spatial and spatio-temporal triangulation  to share boundaries with each other, as is not uncommon 
(see, e.g.,~\cite{es-97}). 

Our spa\-tio-tem\-po\-ral triangulation is also \emph{invariant under time-dependent affinities}.
In the area of spatial  
database research,
much attention has been payed to affine
invariance of both data description and manipulation 
techniques and queries~\cite{ghk-01,gvv-jcss,pods-94}. 
The main idea of working in an  affine invariant way is to obtain
methods and techniques that are not affected by affine transformations of the ambient space in which the data is situated. 
This means that a particular choice of origin or some particular, possibly artificial, choice of unit of measure (e.g.,
inches, centimeters, ...) and direction of coordinate axis has no effect on the final result of the method, technique  or query. This means that an affine-invariant method is robust with respect to a particular choice of measuring data.

Also in other areas, invariance under affinities is often relevant.  In computer vision, the
so-called {\em weak perspective assumption}~\cite{weakpersp} is widely adopted. 
This 
assumption says that when an object is repeatedly photographed from different viewpoints, and the object
is relatively far away form the camera, that all pictures of the object are affine images of each
other, i.e., all images are equal up to an affinity of the photographic plane. We generalize this assumption for \st objects as follows. If a \st event is filmed by
two moving observers, relatively far away from the event, then both films will be the same up to a
time-dependent affinity of the plane of the pellicle. 
For each time moment, another affinity maps the snapshots of the different
movies onto each other.

The weak perspective assumption has  necessitated affine-in\-va\-riant similarity measures
between pairs of pictures~\cite{sym-diff,pattern-hausdorff,geom-hashing}. 
Also, in computer graphics,
affine-in\-va\-riant norms and triangulations have been studied~\cite{nielson}. 
In the field of spatial and spa\-tio-tem\-po\-ral constraint databases~\cite{cdbook,reveszbook},
affine-in\-va\-riant query languages~\cite{ghk-01,gvv-jcss,pods-94} 
have been proposed.
For spatial data,
there exist several triangulation algorithms, but, apart from the triangulation of
Nielson~\cite{nielson}, their output is not affine-invariant. The method proposed by Nielson to
triangulate a set of points in an affine-invariant way computes an affine-invariant norm using the
coordinate information of all points, and then uses this norm in the triangulation algorithm. 
We develop a spatial triangulation algorithm that is more intuitive, that is efficiently computable
and that naturally extends to a \st triangulation algorithm.

The main contribution of this paper is an affine-invariant  time-dependent triangulation algorithm that produces a unique  and
affine-invariant triangulation of a spatio-temporal object given as a   
geometric data object.
As mentioned before, a geometric input object for this triangulation algorithm,  
consists of $m$ atomic geometric objects, given by a triangle, a time interval and a 
time-dependent transformation function that we assume to be given as a fraction of polynomial functions of degree at most $d$. 
We show that our triangulation algorithm runs in polynomial time in the size of the input, measured by $m$ and $d$.
 The worst-case time complexity is of order $z(d, \epsilon)dm^5\log m$, where 
 $z(d,\epsilon)$ is the complexity  of finding all roots of
an univariate polynomial of degree $d$, with accuracy $\epsilon$. The  
 maximal number of
atomic objects in the resulting triangulation is of order $m^5d^6$. For static spatial data the time
complexity of triangulating is of order $m^2\log m$ and the number of returned triangles is of order
$m^2.$  These results are summarized in Table~\ref{taboverview}.

\begin{table}
\centerline{
\begin{tabular}{|c|c|c|}\hline
  & Time Complexity & Output complexity \\ \hline\hline Spatial data &  $O(m^2\log m)$& $O(m^2)$ \\
  Spatio-temporal data &   $O(z(d, \epsilon)dm^5\log m)$& $O(m^5d)$ 
  \\\hline
\end{tabular}
} 
\caption{Summary of the complexity results.}\label{taboverview}
\end{table}

We remark that such triangulations could also be computed via general purpose cell
decomposition algorithms, most notably cylindrical algebraic decomposition~\cite{collins}. 
These algorithms are not affine-invariant, however, and are therefore not directly
suitable for the computational task that  we consider.

In this paper, we also show some applications of the proposed triangulation and show that, when computed in a
pre\-pro\-ces\-sing stage, it facilitates the computation of certain types of queries and operations.

The outline of this paper is as follows. In Section~\ref{sec-defs}, we explain the geometric data
model and define spatial and spa\-tio-tem\-po\-ral triangulations. We introduce an affine invariant
spatial triangulation method in Section~\ref{sec-triang-s}. Afterwards, in
Section~\ref{sec-triang-st}, we describe a novel affine-in\-va\-riant triangulation of
spa\-tio-tem\-po\-ral data. We describe the algorithm in detail and give and prove 
some properties. Then, we
give some possible applications of the triangulation in Section~\ref{sec-app} and we end with some
concluding remarks in Section~\ref{sec-conclusion}.

\section{Preliminaries and Definitions}\label{sec-defs}

We denote the set of real numbers by $\R$ and the two-dimensional real space  by $\R^2$. The space
containing moving \ndim{2} objects will be denoted $\RnR{2}$. We will use $x$ and $y$ (with or
without subscripts) to denote spatial variables and $t$ (with or without subscripts) to denote time
variables. The letter $T$ (with or without subscripts) will be used to
refer to triangles, which we
assume to be represented by triples of pairs of points in $\Rn{2}$.

In this section, we 
first give the definition of a spatial, a temporal and a spa\-tio-tem\-po\-ral object. Next, we
come back to the need of a normal form. Finally, we define affine triangulations of spatial and of
spa\-tio-tem\-po\-ral data.

\subsection{Spatio-temporal Data in the Geometric Data Model}\label{st-defs}
In this section, we describe the geometric data model as introduced by Chomicki and
Revesz~\cite{cr-99}, in which spatio-temporal data are modeled by geometric objects that in turn  are finite collections of  atomic (geometric)
objects. First, we define temporal, spatial and spatio-temporal data objects. In this definition we work with semi-algebraic sets because these are infinite sets that allow a 
effective finite description by means of polynomial equalities and inequalities.
More formally, a se\-mi-al\-ge\-braic set in $\R^d$ is
a Boolean combination of sets of the form $\{(x_1,x_2,\ldots,x_d)\in \R^d\mid
p(x_1,x_2,\ldots,x_d)>0\}$, where $p$ is a polynomial with integer coefficients in the real
variables $x_1$, $x_2$, ..., $x_d$. Properties of se\-mi-al\-ge\-braic sets are well
known~\cite{bcr}.

\begin{definition}\rm\rm\label{defobj}
A \emph{temporal object} is a se\-mi-al\-ge\-braic subset of $\R$, a \emph{spatial object} is a se\-mi-al\-ge\-braic subset of
$\R^2$ and a {\em spa\-tio-tem\-po\-ral object\/} is a se\-mi-al\-ge\-braic subset of
$\RnR{2}$.\qed
\end{definition}

With the \emph{time domain} of a spa\-tio-tem\-po\-ral object, we mean its projection on the time
axis, i.e., on the third coordinate of $\RnR{2}$. 
It is a well-known property of se\-mi-al\-ge\-braic sets, that this projection is a semi-algebraic set and can therefore be considered a temporal
object~\cite{bcr}.
\begin{figure}[h]
 \centerline{\includegraphics[width=150pt, height=90pt]{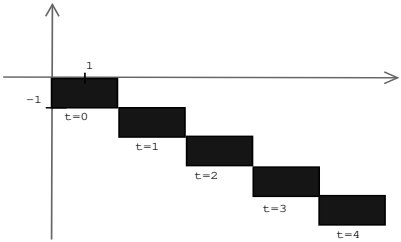}}
  \caption{An example of a  spa\-tio-tem\-po\-ral object.}
  \label{fig2}
\end{figure}

\begin{example}\rm\rm\label{stobject}\rm 
The interval $[0,4]$ and the finite set $\{0,\ab 1,\ab 2,\ab 3,\ab 4\}$ are examples of temporal
objects. The unit circle in the plane is a spatial object, since it can be represented by the
polynomial inequalities $\lnot (1-x^2-y^2>0) \land \lnot (x^2+y^2-1>0)$, usually abbreviated by the
formula $x^2+y^2=1$. The set $\{(x,y;t)\in \R^2\times \R \mid x \geq 2t \land x \leq 2+2t \land y
\geq -t-1 \land y \leq -t \land t \geq 0 \land t \leq 4\}$ is a spa\-tio-tem\-po\-ral object and it
represents rectangle that is translated at constant speed during the time interval $[0,4]$. At each moment $t$ in this
interval it has corner points $(2t,-t)$, $(2+2t, -t)$, $(2+2t, -t-1)$ and $(2t, -t-1)$, as
illustrated in Figure~\ref{fig2}.\qed
\end{example}

In \emph{the geometric data model}~\cite{amai03,cr-99}, spa\-tio-tem\-po\-ral objects are finitely
represented by geometric objects, which in turn are finite collections of atomic geometric objects. An
atomic geometric object is given by its spatial reference object (which determines its shape), a
time interval (which specifies its lifetime) and a transformation function (which determines the
movement of the object during the time interval).

Several classes of geometric objects were introduced, depending on the types of spatial reference
objects and transformations~\cite{amai03,cr-99}.  In this article, we consider \emph{spa\-tio-\-temporal
objects that can be represented as finite unions of triangles moved by time-depen\-dent affine
transformations} (see also Definition~\ref{def-atomic-obj}). Other combinations have been
studied~\cite{amai03} in which triangles, rectangles or polygons
are transformed by time-dependent translations, scalings or affinities, that, in turn, are
given by linear, polynomial or rational functions of time.
 The  class of geometric objects that we consider is not
only the most general of the classes that were previously studied, it is also one of the few
classes that have the desirable property of being closed under the set operations union,
intersection and difference~\cite{amai03}. In Section~\ref{sec-triang-st}, we will rely on this
closure property.

 \begin{definition}\rm\rm\rm
 \label{def-atomic-obj}
An \emph{atomic geometric object\/} $\cali{O}$ is a triple $(T,I,f)$, where
\par\noindent $\bullet$ $T\subset \R^2$ is the {\em spatial reference object} of $\cali{O}$, which is a (filled) \
triangle with corner points that have rational or algebraic coefficients\footnote{For technical reasons, we allow a triangle to degenerate into a line segment or a
point.};

\par\noindent $\bullet$ $I\subset \R$ is the {\em time domain\/} (a point or an interval) of $\cali{O}$; and

\par\noindent $\bullet$ $f:\R^2\times \R \rightarrow \R^2$ is the {\em transformation function} of $\cali{O}$, which is a time-dependent affinity of the form
$$(x,y;t)\mapsto \left(\!\begin{array}{@{}cc@{}}a_{11}(t) & a_{12}(t)\\  a_{21}(t) & a_{22}(t)
\\\end{array}\!\right) \cdot \left(\!\begin{array}{@{}c@{}}x
\\ y  \end{array}\!\right) + \left(\!\begin{array}{@{}c@{}}b_1(t) \\b_2(t)
 \end{array}\!\right),$$
where $a_{ij}(t)$ and $b_i(t)$ are rational functions of $t$ (i.e., of the form $p_1(t)/p_2(t)$,
with $p_1$ and $p_2$ polynomials in the variable $t$ with rational coefficients) and  the determinant of the
matrix of the $a_{ij}$'s differs from zero for all $t$ in $I$. \qed
\end{definition}

We remark that this definition guarantees that there is a finite representation of atomic geometric
objects by means of the polynomial constraint description of the time-interval,  (the cornerpoints
of) the reference triangle and the coefficients of the transformation matrices. 
An atomic geometric
object $\cali{O}=(T,I,f)$ finitely represents the spa\-tio-tem\-po\-ral object
$$\{(x,y;t)\in \R^2\times \R\mid \ab t\in I\ab \land (\exists x')\ab(\exists y')((x', y')\in
T\land  (x, y)=f(x', y'; t))\}, $$ which we denote as $st(\cali{O})$.
Atomic geometric objects can be combined to more complex geometric objects.

\begin{definition}\rm\rm\label{geom-obj}
A {\em geometric object\/}   is a set $\{\cali{O}_1,\ab \ldots, \ab \cali{O}_n\}$ of atomic
geometric objects. It represents the spa\-tio-tem\-po\-ral object $\cup_{i=1}^n st(\cali{O}_i).$
\qed
\end{definition}

By definition~\ref{geom-obj},  the atomic geometric objects that compose a geometric object are allowed  to
overlap in time as well in space. This is a natural definition, but we will see in
Section~\ref{motivation}, that this flexibility in design leads to expensive computations when we want to query spatio-temporal objects represented this way.

We define the {\em time domain} of a geometric object $\{\cali{O}_1,\ab \ldots, \ab \cali{O}_n\}$
to be the smallest time interval that contains all the time intervals $I_i$ of the atomic geometric
objects $\cali{O}_i$ (this is  the convex closure of these time intervals, denoted by
$\overline{\bigcup}_{i=1}^n I_i $).

Remark that a spa\-tio-tem\-po\-ral object is emp\-ty outside the
time domain of the geometric object that defines it. Also, within
the time domain, a spa\-tio-tem\-po\-ral object is empty at any
moment when no atomic object exists.

\begin{example}\rm\rm\label{exampletijdsdomain}
The spa\-tio-tem\-po\-ral object of Example~\ref{stobject} can be represented by the geometric
object $\{\cali{O}_1, \cali{O}_2\}$, where $\cali{O}_1$ is represented by ($T_1$, $[0,4]$, $f$) and
$\cali{O}_2$ equals ($T_2$, $[0,4]$, $f$), with $T_1$ the triangle with corner points $(0,0)$,
$(2,0)$, $(2,-1)$, $T_2$ the triangle with corner points $(0,0)$, $(0,-1)$, $(2,-1)$, and $f$ the
transformation $(x,y;t)$ $\mapsto$ $ (x+2t+2,y-t-1).$ \qed
\end{example}

\begin{figure}
  \centerline{\includegraphics[width=200pt, height = 70pt]{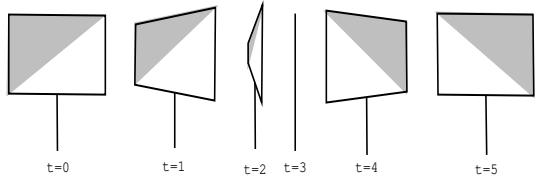}}
  \caption{A spa\-tio-tem\-po\-ral object (a traffic sign) shown at six moments.}\label{figsign}
\end{figure}

\begin{example}\rm\rm\label{ex3}
Figure~\ref{figsign} shows a traffic sign at six moments (seen by an observer walking around it).
This observation can be described by seven atomic geometric objects. During the interval $[0,3[$,
there exist three atomic objects, two triangles, and one line segment. At the time instant $t=3$
there exists one atomic object that represents the shape of a line. During the interval \ $]3,5]$
there exist three atomic objects, two triangles, and a line segment.\qed
\end{example}

To end this subsection, we define the {\em snapshot} of a spa\-tio-tem\-po\-ral object at a certain
moment in time. Snapshots are spatial objects that show what a spa\-tio-tem\-po\-ral object looks
like at a certain moment.

\begin{definition}\rm\rm
Let $\cali{O}$ be an atomic object. Let $\tau_0$ be a time moment in the time domain of $\cali{O}$.
The \emph{snapshot of $\cali{O}$ at time $\tau_0$}, denoted $\cali{O}^{\tau_0}$, is the
intersection of the spa\-tio-tem\-po\-ral object $st(\cali{O})$ with the plane in $\RnR{2}$ defined
by $t = \tau_0$, i.e., the plane $\R^2\times \{\tau_0\}$.

Let $\{\cali{O}_1,\ab \ldots, \ab \cali{O}_n\}$  be a geometric
object. The {\em snapshot of $\{\cali{O}_1,\ab \ldots, \ab
\cali{O}_n\}$ at time $\tau_0$}, denoted $\{\cali{O}_1,\ab
\ldots,\ab \cali{O}_n\}^{\tau_0}$, is the union of the snapshots
at $\tau_0$ of all atomic objects that compose
$\{\cali{O}_1,\ldots, \cali{O}_n\}$, i.e., $\cup_{i=1}^n
\cali{O}_i^{\tau_0}$. \qed
\end{definition}

As explained in Example~\ref{ex3}, Figure~\ref{figsign} shows six snapshots of a geometric object representing a traffic sign seen  by
a moving observer.

\subsection{The Benefits of a Normal Form}\label{motivation}
As remarked after Definition~\ref{geom-obj}, the atomic objects that compose a geometric object may
overlap both in space as in time.
As a consequence, it is impossible to answer some very basic
queries about a geometric object without a lot of computations.

To know the time domain of a spatio-temporal object, for example,
one needs to check all atomic objects that describe it, sort the begin and end points of their time domains, and derive the union of all time domains.

Also, there might be atomic objects that do not
contribute at all to the shape of the spatio-temporal
object as they are entirely overlapped by other
atomic objects. These objects are taken along
unnecessarily in computations. Furthermore,
two geometric objects that represent the same
spatio-temporal object can have a totally different
representation by means of atomic objects. It
is computationally expensive to derive from their
different representations that they are actually the same.

These drawbacks can be solved by introducing a \emph{normal form} for geometric objects, that
makes their structure more transparent. This normal form should have the property that it is the
same for all geometric objects that represent the same spatio-temporal object, independent of their
initial representation by means of atomic objects. In this paper, we add the requirement that this
normal form should be invariant under affinities. 
If two geometric objects are the same up to time-dependent affinities, 
we also want their normal form representation to be the same up to these affinities.

\subsection{Affine Triangulation Methods}

We end this section with the definition of affine spatial  and \st triangulation methods.

 \begin{definition}\rm\rm[\textbf{Spatial and spatio-temporal triangulation}]\label{def-triangulation}
Let $\{{\cal O}_1,\ab \ldots, \ab {\cal O}_n\}$  be a geometric object and $\tau_0$ be a time
moment in the time domain of \{${\cal O}_1, \ab\ldots, \ab{\cal O}_n$\}.

\par\noindent $\bullet$ A collection
of triangles\footnote{Remark that we consider filled triangles and we allow a triangle to
degenerate into a closed line segment or a point.} $\{\trc{1}, \trc{2}, \ldots, \trc{m}\}$ in
$\Rn{2}$ is a \emph{triangulation} of the snapshot $\{{\cal O}_1, \ab\ldots, \ab{\cal
O}_n\}^{\tau_0}$ if the interiors\footnote{We define the \emph{interior} as follows: the interior
of a triangle is its topological interior; the interior of a line segment is the segment without
endpoints; and the interior of a point is the point itself.} of different $T_i$ are disjoint and
the union $\cup_{i=1}^m \trc{i}$ equals $\{{\cal O}_1,\ab \ldots, \ab {\cal O}_n\}^{\tau_0}$.

\medskip
\par\noindent $\bullet$ A geometric object $\{{\cal T}_1,\ldots, {\cal T}_m\}$   is a \emph{triangulation of a geometric
object $\{{\cal O}_1,\ab\ldots, \ab{\cal O}_n\}$} if for each $\tau_0$ in the time domain of
$\{{\cal O}_1,\ab \ldots, \ab {\cal O}_n\}$, $\{{\cal T}_1,\ab \ldots, \ab {\cal T}_m\}^{\tau_0}$
is a triangulation of the snapshot $\{{\cal O}_1,\ab \ldots, \ab {\cal O}_n\}^{\tau_0}$ 
and if furthermore
$st(\{{\cal O}_1,\ldots, {\cal O}_n\})$ $=st(\{{\cal T}_1,\ldots, {\cal T}_m\})$.\qed
\end{definition}

We remark that in the second part of Definition~\ref{def-triangulation}, at each moment $\tau_0$ in the time domain of
$\{{\cal O}_1,\ab \ldots, \ab {\cal O}_n\}$,
 ${\cal T}_i^{\tau_0}$ may  be empty (\ie, $\tau_0$ may be outside the time domain of
${\cal T}_i$).

\begin{figure}
  \centerline{\includegraphics[width=250pt, height = 125pt]{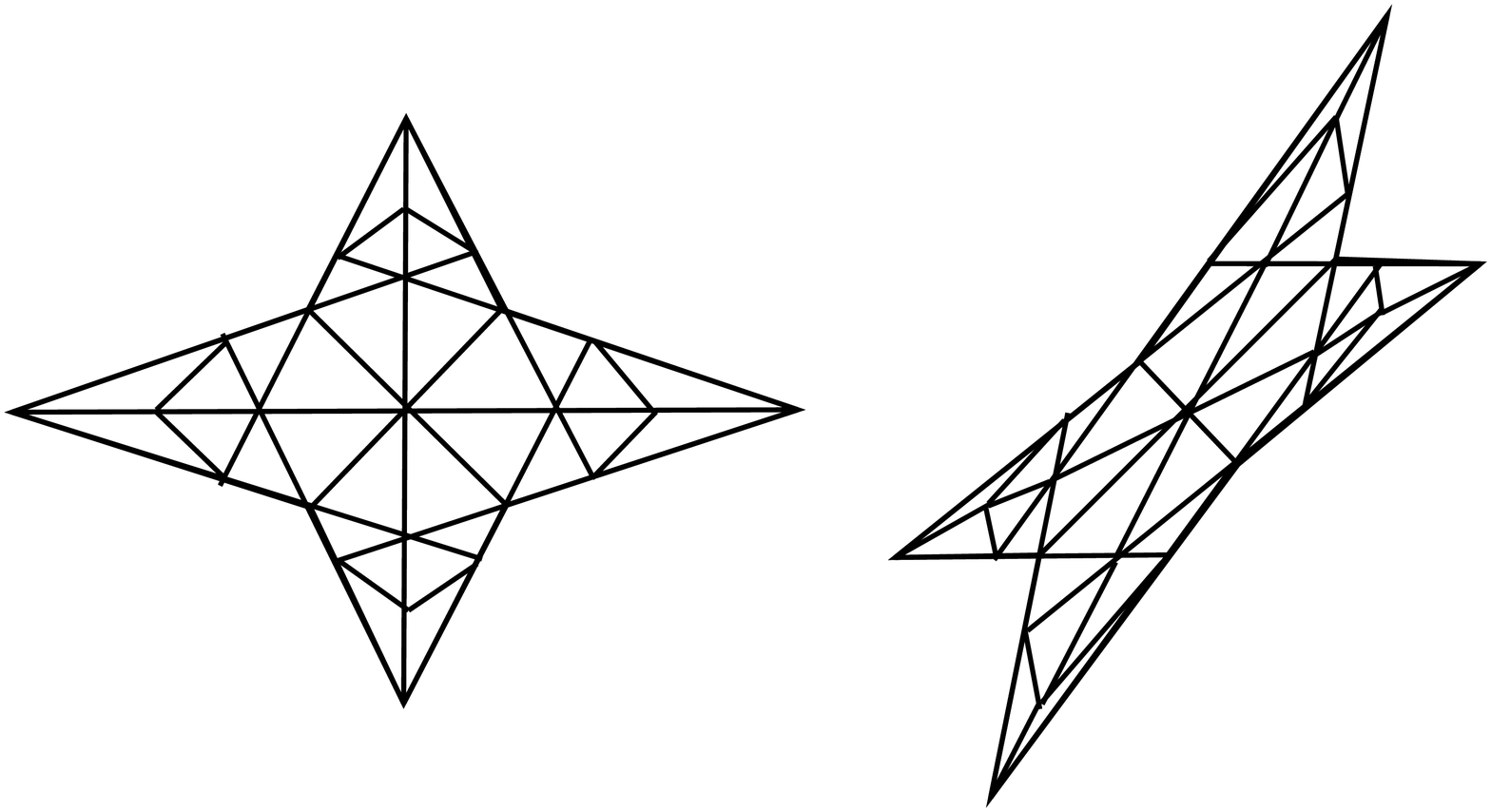}}
  \caption{The triangulations of a snapshot (left) and of an affine transformation of the snapshot (right).}\label{fig-triangulation-stars}
\end{figure}

In Figure~\ref{fig-triangulation-stars}, two stars with their respective triangulations are shown.
Note that, although triangulations of spatial sets intuitively are \emph{partitions} of such sets
into triangles,  they are not partitions in the mathematical sense. Indeed, the elements of the
\emph{partition} may have common boundaries. For spatial data, it is customary to allow the
elements of a partition to share boundaries (see for example~\cite{es-97}).

A \emph{spatial triangulation method} is a procedure that on input (some representation of)  a
snapshot of a \st object produces (some representation of) a triangulation of this snapshot. A
\emph{\sti triangulation method} is a procedure that on input (some representation by means of geometric objects of) a spatio-temporal
object, produces (some representation by means of geometric objects of) a triangulation of this spatio-temporal 
object.

Next, we define what it means for such methods to be affine-in\-va\-riant.

\begin{definition}\rm\rm[\textbf{Affine-invariant triangulation methods}] \label{def-affine-invar-triang}
A spatial triangulation  method ${\cal T}_S$ is called \emph{affine invariant} if and only if for
any two snapshots $A$ and $B$, for which there is an affinity $\alpha :\Rn{2}\rightarrow \Rn{2}$
such that $\alpha(A)=B$, also $\alpha({\cal T}_S(A)) ={\cal T}_S(B)$.

A spa\-tio-tem\-po\-ral triangulation method ${\cal T}_{ST}$ is called \emph{af\-fi\-ne invariant}
if and only if for any  geometric objects $\{{\cal O}_1,\ab \ldots, \ab {\cal O}_n\}$ and $\{{\cal
O}'_1,\ldots, {\cal O}'_m\}$ for which for each moment $\tau_0$ in their time domains, there is an
affinity $\alpha_{\tau_0} :\Rn{2}\rightarrow \Rn{2}$ such that if $\alpha_{\tau_0}(\{{\cal
O}_1,\ldots, {\cal O}_n\}^{\tau_0})=\{{\cal O}'_1,\ldots, {\cal O}'_m\}^{\tau_0},$  also
$\alpha_{\tau_0} ({\cal T}_{ST}(\{{\cal O}_1,\ab \ldots, \ab {\cal O}_n\})^{\tau_0})$  $= {\cal
T}_{ST}(\{{\cal O}'_1,\ab \ldots, \ab {\cal O}'_m\})^{\tau_0}.$\qed
\end{definition}

\begin{figure}
  \centerline{\includegraphics[width=250pt, height = 75pt]{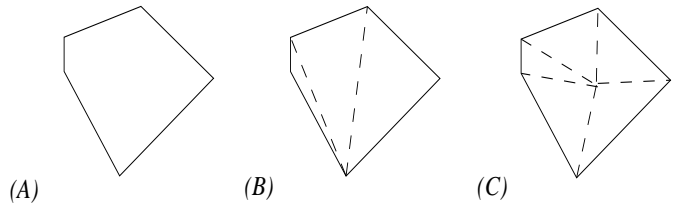}}
  \caption{Two different triangulations ($(B)$ and $(C)$) of a convex polygon ($(A)$).}\label{fig-severaltriangs}
\end{figure}

\begin{example}\rm\rm\rm\label{ex-def-affine-invar-triang}
Given a convex polygon as shown in $(A)$ of Figure~\ref{fig-severaltriangs}. A spatial
triangulation method that takes the leftmost of the corner points with the smallest $y$-coordinates
of the polygon and connects it with all other corner points, is not affine invariant. It is not
difficult to see that, when an affine transformation is applied to the polygon, another point may
become the leftmost lowest corner point. Part $(B)$ of Figure~\ref{fig-severaltriangs} shows the
result of applying this triangulation method to the convex polygon shown in $(A)$.

A triangulation method that computes the barycenter of a convex polygon and connects it with all
corner points is affine-invariant. An illustration the output of this method applied to the polygon shown in $(A)$ is shown in $(C)$ of
Figure~\ref{fig-severaltriangs}. \qed\end{example}

We now propose an affine-invariant spatial triangulation method for spatial figures that are
snapshots of geometric objects, or, that can be represented as finite sets of triangles.

\section{An Affine-in\-va\-riant Spatial Triangulation Method}\label{sec-triang-s}

We next propose an affine-invariant  triangulation method. Later on, in
Section~\ref{sec-triang-st}, we will use the technique proposed here to construct a \st
triangulation algorithm. We first explain the intuition behind the triangulation method, and then
give the details in Algorithm~\ref{algo-triang-snapshot}. We illustrate the algorithm with an
example, prove its correctness and end with determining the size of the output and the time
complexity of the algorithm.

\medskip

\begin{figure}
  \centerline{\includegraphics[width=300pt]{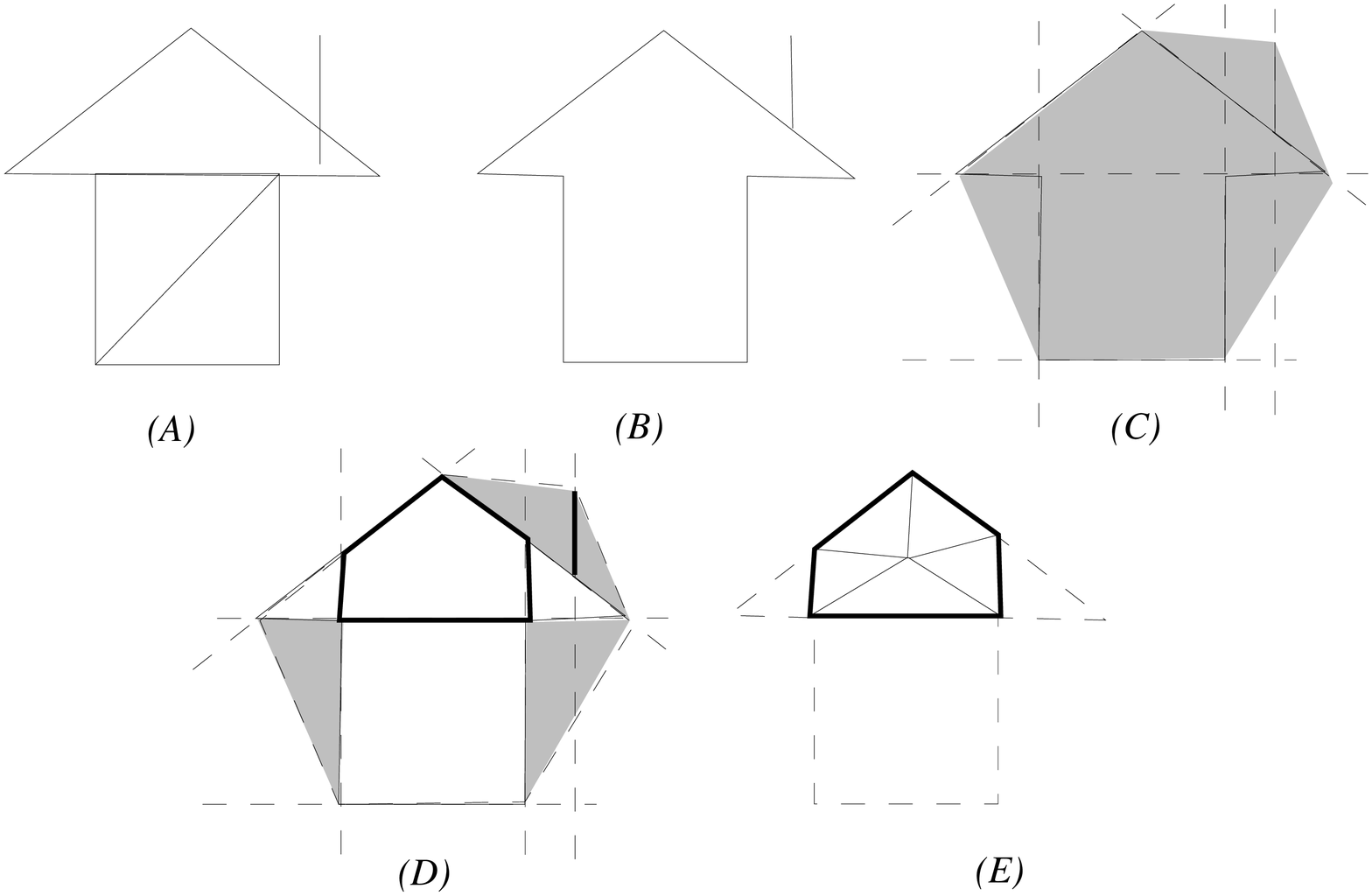}}
  \caption{The several steps in the spatial triangulation algorithm.}\label{fig-intuitiontriang}
\end{figure}

Intuitively, the algorithm is as follows. The input is a snapshot $S$, given as a finite set of
triangles. In Figure~\ref{fig-intuitiontriang} $(A)$, for example, a snapshot of a house-like shape
is given by four triangles. One of those triangles is degenerated into a line segment (representing
the chimney). To make sure that the triangulation is independent of the exact representation of the
snapshot by means of triangles, the boundary of the snapshot, \ie, the boundary of the union of the
triangles composing $S$, is computed. For the snapshot of Figure~\ref{fig-intuitiontriang}, the
boundary is shown in $(B)$. The (triangle degenerated into a) line segment contributes to the
boundary. Therefor, we label it, the reason for this will become clear in a further stage of the
procedure. Also, the (triangles degenerated into) points of the input that are not part of a line
segment or real triangle, \ie, the ones contributing to the boundary, are added to the output
immediately.

 The set of all lines through the edges of the boundary partitions the
plane into a set of open convex polygons, open line segments, open (half-) lines and points. The
(half-) lines and some of the polygons can be unbounded, so we use the convex hull ${\cal CH}(S)$
of the corner points of all triangles in the input as a bounding box. In  $(C)$ of
Figure~\ref{fig-intuitiontriang}, the grey area is the area inside of the convex hull. The
partition of the area inside the convex hull is computed. The points in this partition are not
considered. The points contributing to the boundary were already added to the output in an earlier
stage. For each open line segments, it is checked whether it is part of a labelled line segment of
the input. Recall that only line segments that contribute to the boundary are labelled in an
earlier stage of the algorithm. Only if an open line segment is part of a labelled segment, as is
the case for the one printed in bold in Figure~\ref{fig-intuitiontriang}  $(D)$, its closure
(\ie, a closed line segment) is added to the output. For each open polygon in the partition, we
compute the polygon that is its closure and triangulate this polygon using its center of mass (see
Figure~\ref{fig-intuitiontriang} $(D)$ for a polygon in the partition and $(E)$ for its
triangulation). Some open polygons are only part of the convex hull of $S$, but not of the snapshot
itself. The polygons shaded in grey in  $(D)$ of Figure~\ref{fig-intuitiontriang} are an
example of such polygons.  If a polygon does not belong to $S$, we do not triangulate it. The
triangulations of all other polygons are added to the output. Note that we can decide whether a
polygon belongs to the snapshot by first computing the \emph{planar subdivision} ${\cal U}(S)$
(which we will define next) of the input snapshot and then test for each open polygon whether its
center of mass belongs to the interior of a region or face in the subdivision. We will explain this
in more detail when analyzing the complexity of the algorithm.

\medskip

In the detailed description of the algorithm, we will use some well known techniques. One of those
is the \emph{doubly-connected edge list}~\cite{comp-geom}, used to store \emph{planar
subdivisions}.

\begin{definition}\rm\rm[\textbf{Planar subdivision}]\label{def-planar-sub}
A \emph{planar subdivision} is a subdivision of the plane into labelled regions (or \emph{faces}),
edges and vertices, induced by a plane graph. The \emph{complexity of a subdivision} is the sum of
the number of vertices, the number of edges and the number of faces it consists of.
\qed
\end{definition}

Next, we describe the \emph{doubly-connected edge list}, a data structure to store planar
subdivisions. For this structure, each edge is split into two directed \emph{half-edges}. In
general, a doubly-connected edge list contains a record for each face, edge and vertex of the
planar subdivision.

\begin{definition}\rm\rm[\textbf{Doubly-connected edge list}]\label{def-doubly-conn-edge}
Given a planar subdivision ${\cal S}$. A \emph{doubly-connected edge list for ${\cal S}$}, denoted
\mbox{DCEL}($\cal S$), is a structure containing a record for each face, edge and vertex of the
subdivision.
 These records store the following geometric and topological information:
\begin{enumerate}[(i)]
    \item The \emph{vertex record} of a vertex $\vect{a}$ stores the coordinates of $\vect{a}$ and a pointer to an
    arbitrary half-edge that has $\vect{a}$ as its origin;
    \item The \emph{face record} of a face $f$ stores a pointer to an arbitrary half-edge on its boundary.
    Furthermore, for each hole in $f$, it stores a pointer to an arbitrary half-edge on its
    boundary;
    \item The \emph{half-edge record} of a half-edge $e$ stores five pointers. One to its origin-vertex,
    one to its twin half-edge, one to the face it bounds, and one to the previous and next
    half-edge on the boundary on that face.\qed
\end{enumerate}\end{definition}

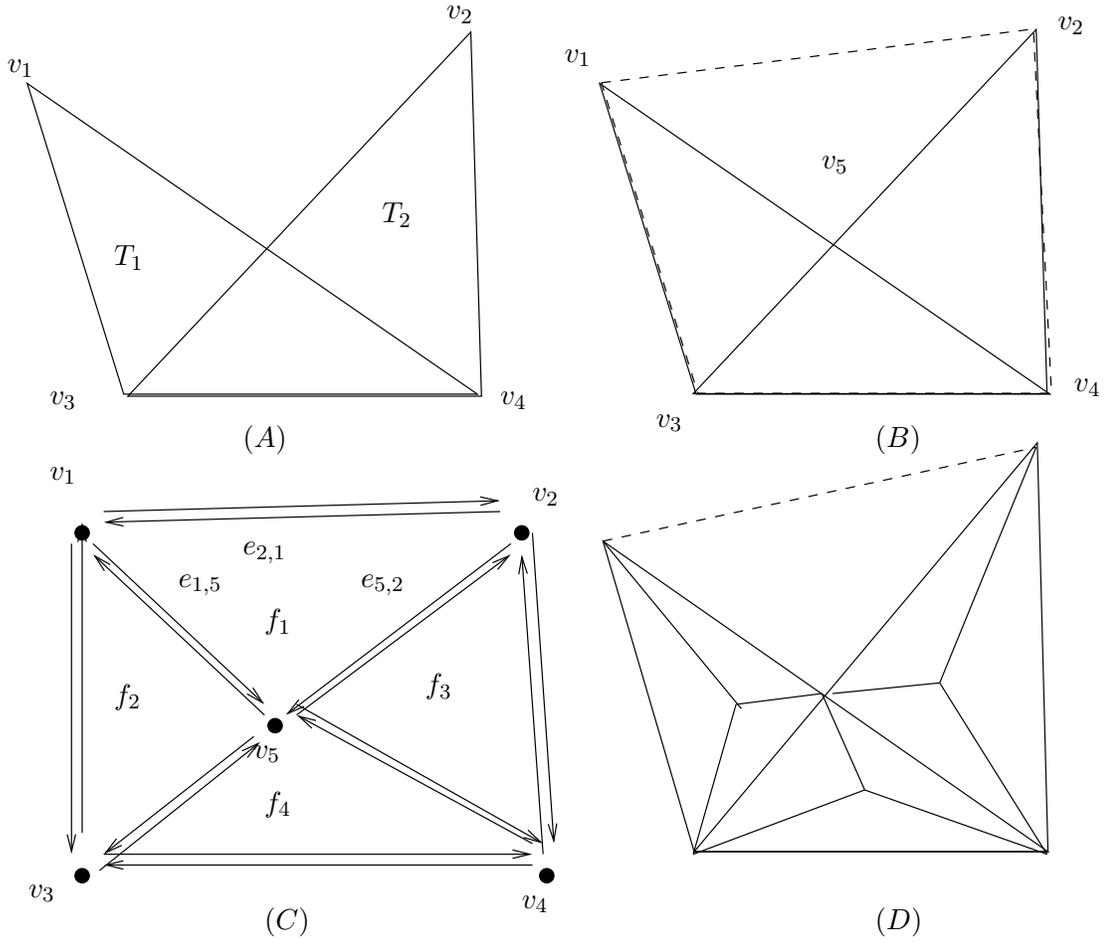
\begin{figure}
\begin{center}
\input{spatialalgo.pstex_t}
\end{center}
  \caption{The different steps of Algorithm~\ref{algo-triang-snapshot} applied to $S = \{T_1, T_2\}$.
  In this example, all  boundary segments of all triangles of $S$ contribute to the boundary of $S$. The line arrangement induced by the carriers of the edges of the input triangles is bounded by the convex hull of the input in $(B)$.
  A doubly-connected edge list is constructed out of the line arrangement, storing its topological structure in $(C)$. Finally, the triangulation is computed in $(D)$.}
  \label{fig-algo-triang-snapshot}
\end{figure}

\begin{example}\rm\rm\rm
Figure~\ref{fig-algo-triang-snapshot} shows a planar subdivision in $(B)$ and 
its topological
structure in $(C)$, that is reflected in the doubly-connected edge list represented in
Table~\ref{table-doubly-triang-snapshot}. \qed\end{example}


Algorithm~\ref{algo-triang-snapshot} (or ${\cal T}_S$) gives the triangulation procedure more
formally. The input of this triangulation algorithm is a snapshot $S$, consisting of a geometric
object which we assume to be given as a finite set of (possibly overlapping and possibly
degenerated) triangles. We further assume that each triangle is represented as a triple of pairs of
coordinates, which are rational numbers.

To shorten and simplify the exposition of Algorithm~\ref{algo-triang-snapshot}, we assume that $S$
is fully two-dimensional, or equivalently, that points and line segments that are not adjacent to a
polygon belonging to $S$ are already in the output. Including their triangulation in the algorithm
would make its description tedious, as we would have to add, and consider, more node and edge
labels.

We use C programming-style notation for pointers to records and elements of records. For example,
Let $\vect{a} = (a_1, a_2)$. In the vertex record $V_a$ of $\vect{a}$, $V_a.x = a_1$ and $V_a.y =
a_2$. Let $e$ be an edge record. The coordinates of the origin $e\rightarrow\text{origin}$ of $e$
are $e\rightarrow\text{origin}\rightarrow x$ and $e\rightarrow\text{origin}\rightarrow y$.

\medskip

\begin{algorithm}\rm\caption{${\cal T}_S$ (Input: $S$ = \{$\trc{1}$, $\trc{2}$, \ldots, $\trc{k}$\}, Output: \{$\trc{1}'$, $\trc{2}'$, \ldots, $\trc{\ell}'$\})}\label{algo-triang-snapshot}
\begin{algorithmic}[1]
    \STATE {\bf Out}:= $\emptyset$.
    \STATE Compute the set ${\cal B}(S)$ containing all line segments, bounding a triangle of the input, that
contribute to the boundary of $S$ (\ie, that contain an edge of the boundary). Meanwhile, construct
the planar subdivision ${\cal U}(S)$ induced by the triangles composing $S$.
    \STATE Compute the convex
hull ${\cal CH}(S)$ of $S$.
    \STATE Construct the doubly connected edge list \mbox{DCEL}($S$),
induced by the planar subdivision defined by the lines through the segments of ${\cal B}(S)$, using
${\cal CH}(S)$ as a bounding box.
    \WHILE {there are any unvisited half-edges in \mbox{DCEL}($\cal
S$) left}
        \STATE Let $e$ be an unvisited edge.
        \STATE $\Sigma_x := 0$, $\Sigma_y  := 0$, count := $0$, Elist := $\emptyset$.
        \WHILE{$e$ is unvisited}
            \STATE Mark $e$ with the label \emph{visited}.
            \STATE Elist $:=$ Elist $\cup\ \{(e\rightarrow\textrm{origin}, e\rightarrow\textrm{next}\rightarrow\textrm{origin})\}$, $\Sigma_x := \Sigma_x + e\rightarrow\textrm{origin}\rightarrow x$, $\Sigma_y := \Sigma_y + e\rightarrow\textrm{origin}\rightarrow
        y$, count := count $+\ 1$.
            \STATE $e := e\rightarrow\textrm{next}$.
        \ENDWHILE
        \STATE $\vect{x} := (\frac{\Sigma_x}{\textrm{count}}, \frac{\Sigma_y}{\textrm{count}})$.
        \IF {the point $\vect{a}$ in $\vect{x}$  belongs to a face of ${\cal U}(S)$}
            \FORALL {elements $(\vect{a}_s, \vect{a}_e)$ of Elist}
            \STATE  {\bf Out} := {\bf Out} $\cup \{\trc{\vect{a}\vect{a}_s\vect{a}_e}\}$, where $\trc{\vect{a}\vect{a}_s\vect{a}_e}$ is the (closed) triangle with corner points $\vect{a}$, $\vect{a}_s$ and
           $\vect{a}_e$.
            \ENDFOR
        \ENDIF
    \ENDWHILE
    \STATE {\bf return} {\bf Out}.
\end{algorithmic}
\end{algorithm}

\medskip

\ignore{In the following, we will denote the result of applying algorithm ${\cal T}_S$ to a
snapshot $A$ by ${\cal T}_S(A)$.}

Before proving the correctness of the algorithm and determining the size of the output and the time
complexity of the algorithm, we give an example.

\medskip

\begin{example}\rm\rm\rm\label{ex-algo-triang-snapshot}
Let $S$ be the set $\{\trc{1}, \trc{2}\}$, where $\trc{1}$ is the triangle with corner points $v_1$,
$v_3$ and $v_4$, and $\trc{2}$ the triangle with corner points $v_2$, $v_3$ and $v_4$, as depicted in
Figure~\ref{fig-algo-triang-snapshot}. The doubly-connected edge list corresponding to  $(C)$
is shown in Table~\ref{table-doubly-triang-snapshot}. We omitted the structures for vertices and
faces, as we don't need them for the second part of the algorithm.

After the doubly-connected edge list is constructed, we create and output the triangles. This is
done by visiting all half-edges once. Suppose we start with $e_{2,1}$. The next-pointers lead to
$e_{1,5}$ and $e_{5,2}$. The next pointer of the last one points to $e_{2,1}$, which we visited
already. This means we visited all edges of one polygon. The center of mass can now be computed and
the triangles added to the output. This is done for all polygons that are part of the input. In
this example, the polygon with corner points $v_1$, $v_5$ and $v_2$ will not be triangulated, as it is
not part of the input. The algorithm stops when there are no more unvisited edges left. \qed
\end{example}

Note that, as an optimization, we could decide to not triangulate faces that are triangles already.
This does not influence the complexity results, however. Therefor, and also for a shorter and more
clear exposition, we formulated the algorithm in a more general form.

\begin{table}
\centerline{
\begin{tabular}{|c||c|c|c|c|c|c|}\hline Half-edge &
Origin & Twin & IncidentFace & Next & Prev \\ \hline\hline $e_{1,2}$ & $v_1$ & $e_{2,1}$ & $f_5$&
$e_{2,4}$ & $e_{3,1}$\\\hline $e_{2,1}$ & $v_2$& $e_{1,2}$ & $f_1$& $e_{1,5}$ & $e_{5,2}$
\\\hline $e_{1,3}$ & $v_1$ &$e_{3,1}$ & $f_2$& $e_{3,5}$ & $e_{5,2}$
\\\hline $e_{3,1}$ & $v_3$ & $e_{1,3}$& $f_5$ &$e_{1,2}$ & $e_{4,3}$
\\ \hline$e_{1,5}$ & $v_1$& $e_{5,1}$ &$f_1$& $e_{5,2}$ &$e_{2,1}$  \\\hline
$e_{5,1}$ & $v_5$& $e_{1,5}$ &$f_2$ &$e_{1,3}$ &$e_{3,5}$  \\\hline $e_{2,4}$ & $v_2$& $e_{4,2}$ &
$f_5$&$e_{4,3}$ & $e_{1,2}$
\\\hline $e_{4,2}$& $v_4$ & $e_{2,4}$ & $f_3$& $e_{2,5}$&$e_{5,4}$
\\\hline $e_{2,5}$ & $v_2$ & $e_{5,2}$ &$f_3$& $e_{5,4}$&$e_{4,2}$\\\hline
$e_{5,2}$ & $v_5$ & $e_{2,5}$ &$f_1$&$e_{2,1}$&$e_{1,5}$\\\hline $e_{3,4}$ & $v_3$ & $e_{4,3}$
&$f_4$&$e_{4,5}$&$e_{5,3}$
\\\hline $e_{4,3}$ & $v_4$ & $e_{3,4}$ &$f_5$&$e_{3,1}$&$e_{2,4}$\\\hline
$e_{3,5}$ & $v_3$ & $e_{5,3}$ &$f_2$&$e_{5,1}$&$e_{1,3}$
\\\hline $e_{5,3}$ & $v_5$ & $e_{3,5}$ &$f_4$&$e_{3,4}$&$e_{4,5}$\\
\hline $e_{4,5}$ & $v_4$ & $e_{5,4}$ &$f_4$&$e_{5,3}$&$e_{3,4}$\\
\hline $e_{5,4}$ & $v_5$ & $e_{4,5}$ &$f_3$&$e_{4,2}$&$e_{2,5}$\\
\hline
\end{tabular}}
\caption{The half-edge records of the doubly-connected edge list corresponding to
Figure~\ref{fig-algo-triang-snapshot}.}\label{table-doubly-triang-snapshot}
\end{table}
 
\medskip

We now prove compute the complexity of both the output and execution time of the triangulation
method described in Algorithm~\ref{algo-triang-snapshot} and afterwards show that it is
affine-invariant. First, we show that ${\cal T}_S$ is indeed a triangulation method.

\begin{property}[\textbf{Algorithm~\ref{algo-triang-snapshot} is a triangulation method}]\label{remark-spatial-triang-istriang}\rm
Let $S$ be a snapshot. The output ${\cal T}_S(S)$ of Algorithm~\ref{algo-triang-snapshot} applied
to $S$ is a triangulation of $S$.
\end{property}
\par\noindent{\bf Proof.}
Let the set of triangles $\{\trc{1}, \trc{2}, \ldots, \trc{k}\}$ determine  a snapshot $S$. 
It is easy to see that the output
${\cal T}_S(S)= \{\trc{1}', \trc{2}', \ldots, \trc{\ell}'\}$ of ${\cal T}_S$ is a triangulation. By
construction, ${\cal T}_ S(S)$ is a set of triangles that either have no intersection, or share a
corner point or bounding segment. It is clear from the algorithm that $\bigcup_{i = 1^k}\trc{i} =
\bigcup_{j = 1^\ell}\trc{j}'$, because each triangle in ${\cal T}_S(S)$ is tested for membership of
$S$. We are also sure that $S$ is covered by the output, because initially, the convex hull of $S$ is triangulated,
which contains $S$.\qed

\medskip

\begin{property}[\textbf{Quadratic output complexity}]\rm\label{prop-complex-output--triang-snapshot}
Let a snapshot $S$ be given by the set $ \{\trc{1}, \ab\trc{2}, \ab\ldots, \ab\trc{m}\}$, consisting of  $m$ triangles.
 The triangulation ${\cal T}_S(S)$, where ${\cal T}_S$ is the triangulation method described in Algorithm~\ref{algo-triang-snapshot}, contains $O(m^2)$ triangles.
\end{property}
\par\noindent{\bf Proof.}
It is well-known (see, e.g., \cite{comp-geom-chap8}), that an arrangement of $m$ lines in the plane
results in a subdivision of the plane containing $O(m^2)$ lines, $O(m^2)$ edges and $O(m^2)$ faces.
It follows that the structure \mbox{DCEL}($\cal S$) will contain $O(m^2)$ half-edges, \ie, two
half-edges for each edge in the arrangement. In the worst case scenario, when all faces of the
partition of the bounding box belong to $S$, one triangle is added to the output for each half-edge
in \mbox{DCEL}($\cal S$) (connecting that half-edge with the center of mass of the face it bounds).
Therefor, the output contains $O(m^2)$ triangles. \qed

\medskip
In the following analysis of the running time of Algorithm~\ref{algo-triang-snapshot}, we assume
that triangles are represented as triples of points, and that a point is represented as a pair of
	rational or algebraic  numbers. We further assume that all basic  arithmetic operations on coordinates of points
require constant time.

\begin{property}[\textbf{$O(m^2\log m)$ running time}]\rm\label{prop-complex-time-triang-snapshot}
Let a snapshot $S$ be given by the set $ \{\trc{1}, \ab\trc{2}, \ab\ldots, \ab\trc{m}\}$, consisting of  $m$ triangles.
 The triangulation method ${\cal T}_S$, described in Algorithm~\ref{algo-triang-snapshot}, computes the triangulation ${\cal T}_S(S)$ of $S$ in time $O(m^2\log m)$.
\end{property}
\par\noindent{\bf Proof.}
Let a snapshot $S$ be given by the set $ \{\trc{1}, \ab\trc{2}, \ab\ldots, \ab\trc{m}\}$. Using a plane-sweep
algorithm~\cite{comp-geom-chap2}, we compute both the list of segments contributing to the boundary
of $S$ and the planar subdivision ${\cal U}(S)$ induced by $\bigcup_{i = 1}^m \trc{i}$. This takes
$O(m^2\log m)$, as there are at most $m^2$ intersection points between boundary segments of
triangles of $S$.


The $m$ triangles composing $S$ together have at most $3m$ different corner points. Computing the
convex hull of $m$ points in the plane can be done in time $O(m\log m)$
(see~\cite{comp-geom-chap1}). The same authors propose, in~\cite{comp-geom-chap8}, an algorithm to
compute a doubly-connected edge list, representing an arrangements of $m$ lines, in time $O(m^2)$.
We next show that the changes we make to this algorithm do not influence its running time. So, as
${\cal B}(S)$ contains at most all $3m$ line segments, it induces an arrangement of at most $3m$
lines. Hence, Step 3 of Algorithm~\ref{algo-triang-snapshot} also takes time $O(m^2)$.

We changed the original algorithm~\cite{comp-geom-chap8} for computing the doubly-connected edge
list of an arrangement of lines as follows:
\begin{enumerate}[(i)]
    \item We computed the \emph{convex hull} of the input to serve as a bounding box instead of an
    axis-parallel rectangle containing all intersection points of the arrangement.
    The complexity of computing such an axis-parallel rectangle is higher ($O(m^2)$) than
    that of computing the convex hull ($O(m\log m)$).
    \item The cost of constructing the doubly-connected edge list of the convex hull is $O(m)$, as
    the convex hull contains at most $3m$ corner points and the algorithm for computing it, as described
    in~\cite{comp-geom-chap1}, already outputs the corner
    points of the convex hull in circular order. In the original
    algorithm~\cite{comp-geom-chap8} with an axis-parallel bounding rectangle, computing the doubly-connected
    edge list of this rectangle only takes constant time. This extra time does, however, not affect the overall complexity.
    \item The next step of both algorithms involves finding the intersection points between the lines to
    be inserted and the partial
    arrangement induced by the previously inserted lines. In the original algorithm, this is easier
    only for the intersection of a line with the bounding box. For the intersections with all other
    lines in the arrangement, the cost is the same.
\end{enumerate}
The next part of Algorithm~\ref{algo-triang-snapshot} (starting from Line 5) takes time $O(m^2\log
m)$. Each half-edge of the doubly-connected edge list is visited only once. Also, each half-edge is
only inserted once into the set \emph{Elist}, and consulted only once therein to create a triangle.
As an arrangement of $m$ lines in the plane results in $O(m^2)$ edges, the number of half-edges in
\mbox{DCEL}($\cal S$) also is $O(m^2)$. We can, in time $O(m^2)$, preprocess ${\cal U}(S)$ into a
structure that allows point location in $O(\log m)$ time~\cite{pointlocation}. Therefor, testing
for each of the $O(m^2)$ centers of mass whether they are part of the input takes $O(m^2\log m)$.
We can conclude that all parts of Algorithm~\ref{algo-triang-snapshot} run in time $O(m^2\log
m)$.\qed

\medskip
Table~\ref{table-spatial-complex} summarizes the computational complexity of the various parts of
Algorithm~\ref{algo-triang-snapshot}.

\begin{table}
\centerline{
\begin{tabular}{|c|l|c|}\hline
Line(s) & Step & Time complexity \\ \hline\hline 2 & Compute ${\cal B}(S)$ and ${\cal U}(S)$ & $O(m^2\log m)$ \\
3 & Compute  ${\cal CH}(S)$ & $O(m\log m)$\\ 4 & Compute \mbox{DCEL}($\cal S$) & $O(m^2)$ \\
$5-19$ & Polygon extraction and triangulation & $O(m^2\log m)$
\\\hline\hline & Overall time complexity & $O(m^2\log m)$ \\\hline
\end{tabular}
} \caption{The time complexity of the various parts of Algorithm~\ref{algo-triang-snapshot}, when
the input is a snapshot represented by $n$ triangles.}\label{table-spatial-complex}
\end{table}

\begin{property}[\textbf{${\cal T}_S$ is affine-invariant}]\rm\label{prop-aff-inv-triang-snapshot}
 The triangulation method ${\cal T}_S$ is affine-invariant.
\end{property}
\par\noindent{\bf Proof.} According to the definition of affine-invariance of spatial triangulation methods
(Definition~\ref{def-affine-invar-triang}), we have to prove the following. Let $A$ be a snapshot given by the set of triangles $\{\trc{a,1},
\trc{a,2}, \ldots, \trc{a,k}\}$ and $B$ be a snapshot given by the set of triangles $\{\trc{b,1}, \trc{b,2}, \ldots, \trc{b,\ell}\}$, such that there exists an affinity $\alpha: \Rn{2}
\rightarrow \Rn{2}$ for which $B = \alpha(A)$. Then, for each triangle $T$ of ${\cal T}_S(A)$, it
holds that the triangle $\alpha(T)$ is a triangle of ${\cal T}_S(B)$.

We prove this by going through the steps of the triangulation procedure ${\cal T}_S$. Let $A $ and $B$ be as above. \medskip

The convex hull and boundary of spatial figures are both affine-invariant (more specific, the
boundary is a topological invariant). Intersection points between lines and the order of
intersection points on one line with other lines are affine-invariant (even topological invariant).
The subdivision of the convex hull ${\cal CH}(B)$ of $B$ induced by the arrangement of lines
through the boundary of $B$ is hence the image under $\alpha$ of the subdivision of the convex hull
${\cal CH}(A)$ of $A$ induced by the arrangement of lines through the boundary of $A$. The
doubly-connected edge list only stores topological information about the arrangement of lines, \ie,
which edges are incident to which vertices and faces. Naturally, this information is preserved by
affine transformations. The center of mass of a convex polygon is an affine invariant. Finally, the
fact that a triangle is inside the boundary of the input and the fact that it is not are both
affine-invariant. This completes the proof.\qed

\ignore{ 

 We now define what is means for a snapshot to be in \emph{${\cal T}_S$-normal form}.

\begin{definition}\rm\rm
Let $A $ be a snapshot given as  the set of triangles $\{\trc{1}, \trc{2}, \ldots, \ab \trc{k}\}$.
We say that $A$ (or more precisely, this set of triangles)  is in \emph{${\cal T}_S$-normal form}
if ${\cal T}'_S$ on input $A$ returns $\{\Delta_1, \ldots, \Delta_n\}$ as output.
\end{definition}
}

\medskip

Summarizing this section, we proposed a spatial triangulation method that, given a snapshot
consisting of $m$ triangles, returns an affine-invariant triangulation of this snapshot containing
$O(m^2)$ triangles, in time $O(m^2\log m)$.

We remark here that the idea of using carriers of boundary segments to partition figures was also
used in an algorithm to decompose semi-linear sets by Dumortier, Gyssens, Vandeurzen and Van
Gucht~\cite{luc-pods}. Their algorithm is not affine-invariant, however.


In the next section, we will use the affine triangulation for snapshots to construct a
triangulation of geometric objects.

\section{An Affine-in\-va\-riant Spatio-tem\-poral Triangulation Method}\label{sec-triang-st}

In this section, we present an \st triangulation algorithm that takes as input a geometric object,
\ie, a finite set of atomic geometric objects.
 We will adapt the spatial triangulation method ${\cal T}_S$, described in
Algorithm~\ref{algo-triang-snapshot}, for time-dependent data.

The proposed \st triangulation algorithm ${\cal T}_{ST}$ will have three main construction steps.
First, in the \emph{partitioning step}, the time domain of the geometric object will be partitioned
into a set of points and open time intervals. For each element of this partition, all its snapshots
have an \emph{isomorphic triangulation}, when computed by the method ${\cal T}_S$. We refer to
Definition~\ref{def-triang-iso} below for a formal definition of this isomorphism. Second, in the
\emph{triangulation step}, the \st triangulation is computed for each element in the time
partition, using the fact that all snapshots have \emph{isomorphic triangulations}. Third, in the
\emph{merge step}, we merge objects when possible, to obtain a unique (and minimal) triangulation.

We will start this section by defining isomorphic triangulations. Then we explain the different
steps of the algorithm for computing a \st affine-invariant triangulation of geometric objects
separately. We illustrate the algorithm with an example and end with some properties of the
triangulation.

Intuitively, two snapshots $S_1$ and $S_2$ are called ${\cal T}_S$-isomorphic if the triangles in ${\cal T}_S(S_1)
\cup {\cal T}_S({\cal CH}(S_1)\setminus S_1)$ and ${\cal T}_S(S_2) \cup {\cal T}_S({\cal
CH}(S_2)\setminus S_2)$ have the same (topological) adjacency graph. In particular, if $S_1$ and
$S_2$ are equal up to an affinity of $\Rn{2}$, then they are ${\cal T}_S$-isomorphic.

 \begin{definition}\rm\rm[\textbf{${\cal T}_S$-isomorphic snapshots}]\label{def-triang-iso} Let $S_1$ and $S_2$ be two snapshots of a geometric object.
We say that $S_1$ and $S_2$ are \emph{${\cal T}_S$-isomorphic}, denoted $S_1 \equiv_{{\cal T}_S}
S_2$, if there exists a bijective mapping $h: \Rn{2} \rightarrow \Rn{2}$ with the following
property: A triangle $\trc{} = (\vect{a}_1, \vect{a}_2, \vect{a}_3)$ of ${\cal T}_S(S_1)$ is
incident to the triangles $\trc{1,2}$, $\trc{2,3}$ and $\trc{3,1}$ (where each $\trc{i,((i+1)\mod
3)}$ is either a triangle of ${\cal T}_S(S_1)$ that shares the segment
$\vect{a}_i\vect{a}_{((i+1)\mod 3)}$ with $\trc{}$, a triangle of ${\cal T}_S({\cal
CH}(S_1)\setminus S_1)$ that shares the segment $\vect{a}_i\vect{a}_{((i+1)\mod 3)}$ with $\trc{}$,
or is $\epsilon$, which means that no triangle shares that boundary segment with $\trc{}$) if and
only if, the triangle $h(\trc{}) = \ab (h(\vect{a}_1), \ab h(\vect{a}_2), \ab h(\vect{a}_3))$
belongs to ${\cal T}_S(S_2)$ and is  bounded by $h(\trc{1,2})$, $h(\trc{2,3})$ and $h(\trc{3,1})$.
Moreover, if $\trc{i,((i+1)\mod 3)}$ is a triangle of ${\cal T}_S(S_1)$, then $h(\trc{i,((i+1)\mod
3)})$ is a triangle of ${\cal T}_S(S_2)$ that shares the line segment
$h(\vect{a}_i)h(\vect{a}_{((i+1)\mod 3)})$ with $h(\trc{})$, if $\trc{i,((i+1)\mod 3)}$ is a
triangle of ${\cal T}_S({\cal CH}(S_1)\setminus S_1)$, then $h(\trc{i,((i+1)\mod 3)})$ is a
triangle of ${\cal T}_S({\cal CH}(S_2)\setminus S_2)$ that shares the line segment
$h(\vect{a}_i)h(\vect{a}_{((i+1)\mod 3)})$ with $h(\trc{})$ and if $\trc{i,((i+1)\mod 3)}$ equals
$\epsilon$, then so does $h(\trc{i,((i+1)\mod 3)})$.\qed
\end{definition}

\medskip

\begin{figure}[t]
  \centerline{\includegraphics[width=300pt, height = 100pt]{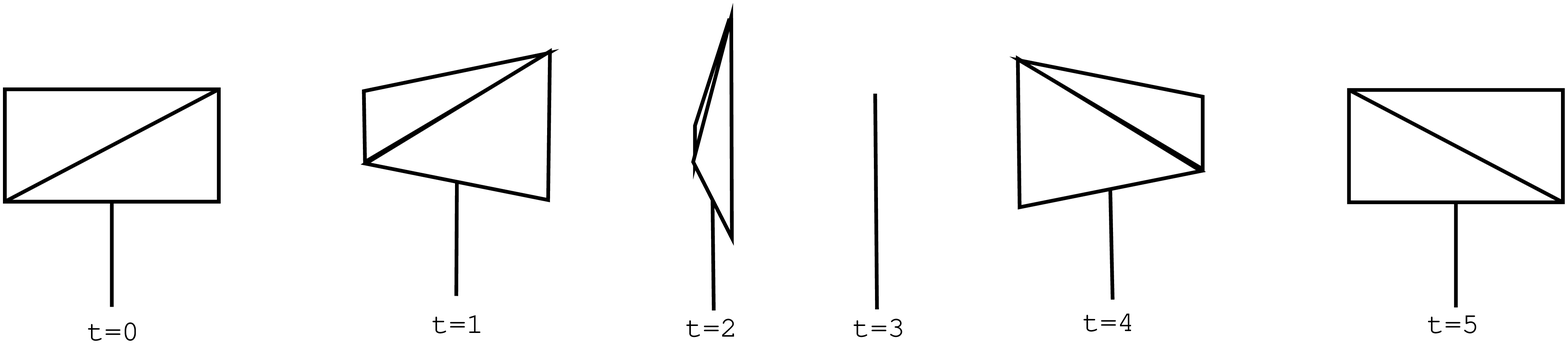}}
  \caption{Snapshots of a traffic sign as seen by an observer circularly moving around it.}
  \label{fig-sign}
\end{figure}

\begin{example}\rm\rm\rm\label{ex-triang-iso}
The triangulations shown in Figure~\ref{fig-triangulation-stars} are ${\cal T}_S$-isomorphic to
each other. In Figure~\ref{fig-sign}, all snapshots shown except the one at time moment $t = 3$ are
${\cal T}_S$-isomorphic. The snapshot at time moment $t = 3$ is clearly not isomorphic to the
others, since it consists only of one line segment.\qed
\end{example}

Remark that for Figure~\ref{fig-triangulation-stars}, the mapping $h$ is an affinity. In
Figure~\ref{fig-sign}, this is not the case.

Now, we introduce a \st triangulation method ${\cal T}_{ST}$ that constructs a time-dependent
affine triangulation of \st objects that are represented by geometric objects. We will explain its
three main steps, \ie, the partitioning step, the triangulation step and the merge step separately
in the next subsections.

We will illustrate each step on the following example.

\begin{figure}
\centerline{\includegraphics[width=350pt, height = 150pt]{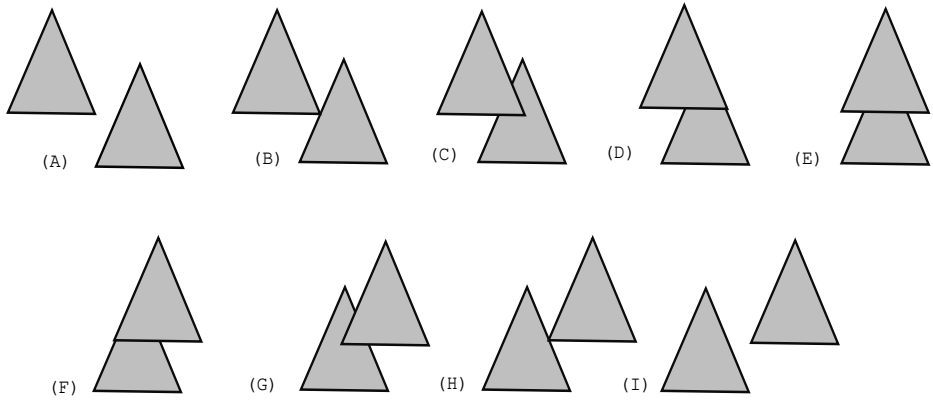}} \caption{The
snapshots at time moments $t = \frac{1}{4}$ {\tt(A)}, $t = \frac{1}{2}$ {\tt(B)}, $t = 1$ {\tt(C)},
$t = \frac{3}{2}$ {\tt(D)}, $t = 2$ {\tt(E)}, $t = \frac{5}{2}$ {\tt(F)}, $t = 3$ {\tt(G)}, $t =
\frac{7}{2}$ {\tt(H)} and $t = 4$ {\tt(I)} of the geometric object of
Example~\ref{ex-running-st}.}\label{fig-triang-st-initial}
\end{figure}

\begin{example}\rm\rm\rm\label{ex-running-st}
Let ${\cal O} = \{{\cal O}_1, {\cal O}_2\}$ be a geometric object, where ${\cal O}_1$ is given as
$(((-1,\ab 0),\ab (1,\ab 0),(0,2)),\ab[0,4], \ab Id)$ and ${\cal O}_2$ is given as $ (((-3,\ab
1),\ab (-1,\ab 1),\ab (-2,\ab 3)),\ab [0,4], f)$ and $f$ is the mapping given by 
$(x, y, t)\mapsto (x+t, y)$. Figure~\ref{fig-triang-st-initial} shows the snapshots of ${\cal
O}$ at time moments $t = \frac{1}{4}$ {\tt(A)}, $t = \frac{1}{2}$ {\tt(B)}, $t = 1$ {\tt(C)}, $t =
\frac{3}{2}$ {\tt(D)}, $t = 2$ {\tt(E)}, $t = \frac{5}{2}$ {\tt(F)}, $t = 3$ {\tt(G)}, $t =
\frac{7}{2}$ {\tt(H)} and $t = 4$ {\tt(I)}.\qed
\end{example}

\medskip

Let ${\cal O}= \{{\cal O}_1 = (S_1, I_1, f_1), {\cal O}_2 = (S_2, I_2, f_2), \ldots, {\cal O}_m =
(S_m, I_m, f_m)\}$ be a geometric object. We assume that the $S_i$ are given as triples of points (\ie, pairs of rational or algebraic numbers),
the $I_i$ as structures containing two rational or algebraic numbers and two flags (indicating whether the interval
is closed on the left or right side) and, finally, the $f_i$ are affinities given by rational functions, i.e.,  fractions of polynomials with integer
coefficients (that we assume to be given in dense or sparse representation), for $i = 1, \ldots, m$.

\subsection{The Partitioning Step.}\label{sec-triang-st-partitionstep} Let ${\cal O}= \{{\cal O}_1
= (S_1, I_1, f_1), {\cal O}_2 = (S_2, I_2, f_2), \ldots, {\cal O}_m = (S_m, I_m, f_m)\}$ be a
geometric object. In the first step of ${\cal T}_{ST}$, the time domain $I$ of ${\cal O}$, \ie, the
convex closure $\overline{\bigcup}_{i=1}^mI_i$ of the union  of all the time domains $I_i (i = 1,
\ldots, m)$ is partitioned in such a way that, for each element of that partition, all its snapshots
are ${\cal T}_S$-isomorphic.

Below, we refer to  the set of lines that intersect the border of $f_i(S_i,\tau)$ in
infinitely many points, the set of \emph{carriers of the snapshot
$f_i(S_i,\tau)$}
and denote it $car(f_i(S_i,\tau))$, ($i = 1, \ldots, m$). 
\medskip

In \cite{amai03}, we defined the \emph{finite time partition} ${\cal P}$ of the time domain of two
atomic objects in such a way that for each element $P$ of ${\cal P}$, the carrier sets of each
snapshot of $P$ are topologically equivalent. This definition can easily be extended to an
arbitrary number of atomic objects. Also the property stating that the finite time partition
exists, still holds in the extended setting.

\begin{definition}\rm\rm[\textbf{Generalized finite time partition}]\label{def-finitetimepartition-general}
\rm We call a {\em finite time partition} of a geometric object ${\cal O} = \{{\cal O}_1, {\cal
O}_2, \ldots, {\cal O}_m\}$ any partition of the interval $I = \overline{\bigcup}_{i=1}^mI_i$ into
a finite number of time intervals $J_1,\ldots, J_k$ such that for any $\tau,\tau'\in J_\ell$ (and
all $1\leq \ell\leq k$), $\bigcup_{i = 1}^m car(f_i(S_i,\tau))$ and $\bigcup_{i = 1}^m
car(f_i(S_i,\tau'))$ are topologically equivalent sets in $\Rn{2}$.\qed
\end{definition}

Here, two subsets $A$ and $B$ of $\Rn{2}$ are called {\em topologically equivalent} when there
exists an orientation-preserving homeomorphism $h$ of $\Rn{2}$ such that $h(A)=B$.

The proof of the following property follows the lines of a proof in~\cite{amai03}.

\begin{property}[\textbf{Existence of a generalized finite time partition}]\label{prop-finitetimepartition-general}
Let ${\cal O}$ $= \{{\cal O}_1, {\cal O}_2, \ldots, {\cal O}_m\}$ be a geometric object.There exists a finite time partition of
${\cal O}$.\qed
\end{property}

\medskip

\medskip

We now proceed with the partitioning step of the \st triangulation algorithm. In this step, a
generalized finite time partition of ${\cal O}$ is computed, using the information of the
\emph{time-dependent} carriers of the atomic objects in ${\cal O}$. Each time an intersection point
between two or more time-dependent carriers starts or ceases to exist, or when intersection points
change order along a line, a new time interval of the partition is started. Given three
\emph{continuously moving} lines, the intersection points of the first line with the two other
lines only change order along the first line, if there exists a moment where all three lines
intersect in one point. Algorithm~\ref{algo-triang-st-split} describes the partitioning step in
detail.

\begin{algorithm}
\caption{\textbf{Partition} (Input: ${\cal O}=\ab\{{\cal O}_1,\ab...,\ab{\cal O}_n\}$, Output:
$\chi = (\tau_1, \tau_2, \ldots, \tau_m)$)}\label{algo-triang-st-split}
\begin{algorithmic}[1]
\STATE Let $\chi = (\tau_1 \leq \tau_2 \leq \ldots \leq \tau_k)$ (with $2 \leq k \leq 2n$) be a sorted
list of time moments that appear either as a begin or endpoint of $I_i$ for any of the objects
${\cal O}_i = (S_i, I_i, f_i)$, $1 \leq i \leq n$.

\STATE ${\cal C} = \emptyset$.

\FORALL{atomic objects ${\cal O}_i = (S_i, I_i, f_i), 1 \leq i \leq n$}
    \STATE Add the new atomic objects $(S_{i,1}, I_i, f_i)$, $(S_{i,2}, I_i, f_i)$ and $(S_{i,3}, I_i, f_i)$ to ${\cal
    C}$, where $S_{i,1}$, $S_{i,2}$ and $S_{i,3}$ are the boundary segments of $S_i$.
\ENDFOR

\FORALL{pairs of objects $(S_{i, {\ell_1}}, I_i, f_i)$ and $(S_{j, {\ell_2}}, I_j, f_j)$ of ${\cal
C}$ $(1 \leq i < j \leq n; 1 \leq \ell_1, \ell_2 \leq 3)$}
    \IF{$I_i \cap I_j \neq \emptyset$}
        \STATE Compute the end points of the intervals during which the intersection of the carriers
        of both time-dependent line segments does exist. Add those such end points that lie within the interval $I_i \cap
    I_j$ to $\chi$, in a sorted way.
    \ENDIF
\ENDFOR

\FORALL{triples of objects $(S_{i, {\ell_1}}, I_i, f_i)$,  $(S_{j, {\ell_2}}, I_j, f_j)$ and
$(S_{k, {\ell_3}}, I_k, f_k)$  of ${\cal C}$ $(1 \leq i < j < k \leq n; 1 \leq \ell_1, \ell_2,
\ell_3 \leq 3)$}
    \IF{$I_i \cap I_j \cap I_k \neq \emptyset$}
        \STATE Compute the end points of the intervals during which the carriers of the three
        time-dependent line segments intersect in one point. Add those such end points that lie within the interval $I_i \cap
    I_j \cap I_k$ to $\chi$, in a sorted way.
    \ENDIF
\ENDFOR

\STATE Return $\chi$.
\end{algorithmic}
\end{algorithm}

We will show later that the result of the generalized finite time partition is a set of intervals
during which all snapshots are ${\cal T}_ S$-isomorphic. This partition is, however, not the
coarsest possible partition having this property, because there might be atomic objects that,
during some time, are completely overlapped by other atomic objects. Therefor, we will later, after
the triangulation step, again merge elements of the generalized finite time partition, whenever
possible.

We illustrate Algorithm~\ref{algo-triang-st-split} on the geometric object of
Example~\ref{ex-running-st}.

\begin{example}\rm\rm\rm\label{ex-running-partition}
Recall from Example~\ref{ex-running-st} that ${\cal O} = \{{\cal O}_1, {\cal O}_2\}$, where ${\cal
O}_1$ is given as $ (((-1,\ab 0),\ab (1,\ab 0),(0,2)),\ab[0,4], \ab Id)$ and ${\cal O}_2$ is given
as $ (((-3,\ab 1),\ab (-1,\ab 1),\ab (-2,\ab 3)),\ab [0,4], f)$ and $f$ is the affinity
mapping triples $(x, y, t)$ to pairs $(x+t, y)$.

We now illustrate the partitioning algorithm on input ${\cal O}$. First, the list $\chi$ will
contain the time moments $0$ and $4$. The list ${\cal C}$ will contain six elements.
Table~\ref{table-ex-triang-st-segments} shows these segments and the formulas describing their
time-dependent carriers. All pairs of segments have an intersection that exists always, except for
the pairs $({\cal O}_{c,2}, {\cal O}_{c,5})$, $({\cal O}_{c,3}, {\cal O}_{c,6})$ and $({\cal
O}_{c,1}, {\cal O}_{c,4})$. The intersections of ${\cal O}_{c,2}$ with ${\cal O}_{c,5}$ and ${\cal
O}_{c,3}$ with ${\cal O}_{c,6}$  exist only at respectively $t = \frac{5}{2}$, $t = \frac{3}{2}$.
The segments ${\cal O}_{c,1}$ and ${\cal O}_{c,4}$ never intersect. Of all possible triples of
carriers, only two triples have a common intersection within the interval $[0,4]$. The carriers of
${\cal O}_{c,2}$, ${\cal O}_{c,4}$ and ${\cal O}_{c, 6}$ intersect at $t = \frac{1}{2}$ and the
carriers of ${\cal O}_{c,3}$, ${\cal O}_{c,4}$ and ${\cal O}_{c, 5}$  intersect at $t =
\frac{7}{2}$. The partitioning step will hence return the list $$\chi = (0, \frac{1}{2},
\frac{3}{2}, \frac{5}{2}, \frac{7}{2}, 4).$$ \qed\end{example}

\begin{table}
\begin{center}
\begin{tabular}{|l|l|}\hline
Element & Carrier \\\hline\hline ${\cal O}_{c,1} = (((-1,0),(1,0)),[0,4],Id)$ & $y = 0$\\\hline
${\cal O}_{c,2} = (((-1,0),(0,2)),[0,4],Id)$ & $y = 2x+2$\\\hline ${\cal O}_{c,3} =
(((0,2),(1,0)),[0,4],Id)$ & $y = -2x+2$\\\hline ${\cal O}_{c,4} = (((-3,1),(-1,1)),[0,4],f)$ & $y =
1$\\\hline ${\cal O}_{c,5} = (((-3,1),(-2,3)),[0,4],f)$ & $y = 2x+7 -2t$\\\hline ${\cal O}_{c,6} =
(((-2,3),(-1,1)),[0,4],f)$ & $y = -2x - 1 +2t$\\\hline
\end{tabular}
\caption{The elements of the list ${\cal C}$ during the execution of the partitioning algorithm
(Algorithm~\ref{algo-triang-st-split}) on the geometric object from
Example~\ref{ex-running-partition}.}\label{table-ex-triang-st-segments}
\end{center}
\end{table}

\medskip

We analyze both the output complexity and sequential time complexity of the partition step. First
remark that the product of $\ell$ univariate polynomials of degree $d$ is a polynomial of degree
$\ell d$. Let the transformation function of an atomic object consists of rational coefficients,
being fractions of polynomials of degree at most $d$. It follows that the time-dependent line
segments and carriers can be defined using fractions of polynomials in $t$ of degree $O(d)$. Also,
the time-dependent intersection point of two such carriers and the time-dependent cross-ratio of an
intersection point compared to two moving end points of a segment, can be defined using fractions
of polynomials in $t$ of degree $O(d)$.

\begin{property}[\textbf{Partition: output complexity}]\rm
\label{prop-complex-output-partition} Given a geometric object ${\cal O}$ $= \{{\cal O}_1 = \ab(S_1,
\ab I_1, \ab f_1), \ab{\cal O}_2 = \ab(S_2, \ab I_2, \ab f_2), \ab\ldots, \ab {\cal O}_n = \ab
(S_n, \ab I_n, \ab f_n)\}$ consisting of $n$ atomic objects. Let $d$ be the maximal degree of any
polynomial in the definition of the transformation functions $f_i, 1 \leq i \leq n$. The procedure
{\bf Partition}, as described in Algorithm~\ref{algo-triang-st-split}, returns a partition of  $I =
\overline{\bigcup}_{i=1}^nI_i$ containing $O(n^3 d)$ elements. \end{property}
\par\noindent{\bf Proof.} It is clear that the list $\chi$ contains $O(n)$ elements after Line 1 of Algorithm~\ref{algo-triang-st-split}. Indeed, at most two elements are added for each
atomic object. The list ${\cal C}$ will contain at most $3n$ elements. For each atomic object with
a reference object that is a ``real'' triangle, $3$ elements will be added to ${\cal C}$. In the
case that one or more corner points coincide, one or two objects will be added to ${\cal C}$.

Now we investigate the number of time moments that will be inserted to $\chi$ while executing the
for-loop starting at Line 6 of Algorithm~\ref{algo-triang-st-split}. The intervals during which the
intersection of two time-dependent carriers exists are computed.  The intersection of two
time-dependent line segments doesn't exist at time moments where the denominator of the rational
function defining it is zero. Because this denominator always is a polynomial $P$ in $t$, it has at
most $\textrm{deg}(P)$ zeroes, where $\textrm{deg}(P)$ denotes the degree of $P$. Accordingly, at
most $\textrm{deg}(P) = O(d)$ elements will be added to $\chi$. Hence, in total, $O(n^2d)$ time
moments are added in this step.

For the intersections of three carriers, a similar reasoning can be used. Hence, during the
execution of the for-loop starting at Line $11$ of Algorithm~\ref{algo-triang-st-split}, $O(n^3 d)$
elements are added to $\chi$.

We can conclude that the list $\chi$ will contain $O(n^3d)$ elements.\qed

\medskip

Now we analyze the time complexity of {\bf Partition}. We first point out that finding all roots of
an univariate polynomial of degree $d$, with accuracy $\epsilon$ can be done in time $O(d^2\log
d\log\log(\frac{1}{\epsilon}))$~\cite{rootfinding}. We will use the abbreviation $z(d, \epsilon)$
for the expression $O(d^2\log d\log\log(\frac{1}{\epsilon}))$. Note also that, although the product
of two polynomials of degree $d$ is a polynomial of degree $2d$, the computation of the product
takes time $O(d^2)$. To keep the proofs of the complexity results as readable as possible, we will
consider the complexity of any manipulation on polynomials (computing zeros, adding or multiplying)
to be $z(d, \epsilon)$, where a precision of $\epsilon$ is obtained.

\begin{property}[\textbf{Partition: computational complexity}]\rm
\label{prop-complex-time-partition} Given a geometric object ${\cal O}= \{{\cal O}_1 = (S_1, I_1,
f_1), {\cal O}_2 = (S_2, I_2, f_2), \ldots, \ab {\cal O}_n = \ab (S_n, I_n, f_n)\}$ consisting of
$n$ atomic objects. Let $d$ be the maximal degree of any polynomial in the definition of the
transformation functions $f_i, 1 \leq i \leq n$ and let $\epsilon$ be the desired precision for
computing the zeros of polynomials. The procedure {\bf Partition}, as described in
Algorithm~\ref{algo-triang-st-split}, returns a partition of  $I = \overline{\bigcup}_{i=1}^nI_i$
in time $O(n^3(z(d,\epsilon) + d\log n))$. \end{property}
\par\noindent{\bf Proof.} Let ${\cal O}= \{{\cal O}_1 = (S_1, I_1,
f_1), {\cal O}_2 = (S_2, I_2, f_2), \ldots, \ab {\cal O}_n = \ab (S_n, I_n, f_n)\}$ be a geometric
object. Let $d$ be the maximum degree of any of the polynomials used in the definition of the
functions $f_i, 1 \leq i \leq n$.

Constructing the initial list $\chi$, on Line 1, takes time $O(n\log n)$ (it is well known that the
inherent complexity of sorting a list of $n$ elements is $O(n\log n)$). Computing the set ${\cal
C}$ can be done in time $O(nd)$: all $n$ elements of ${\cal O}$ are considered, and the time needed
to copy the transformation functions $f_i$ depends on the maximal degree the polynomials defining
them have. Recall that ${\cal C}$ contains at most $3n$ elements.

The first for-loop, starting at Line 6 of Algorithm~\ref{algo-triang-st-split} is executed $O(n^2)$
times. One execution of its body takes $z(d, \epsilon)$. Indeed, computing the formula representing
the time-dependent intersection, checking whether its denominator is always zero and finding the
zeros of the denominator (a polynomial of degree linear in $d$) have all time complexity $z(d,
\epsilon)$. Therefor, the first for-loop takes time $O(n^2z(d, \epsilon))$ in total.

The second for-loop has time complexity $O(n^3z(d, \epsilon))$. The reasoning here is the same as
for the previous for-loop.

Finally, sorting the list $\chi$, which contains $O(n^3d)$ elements at the end, requires
$O(n^3d\log(nd))$.

If we summarize the complexity of all the separate steps, we obtain $O(n^3(z(d,\epsilon) + d\log
n))$. \qed

\medskip

We now proceed with the triangulation step.

\subsection{The Triangulation Step.}\label{sec-triang-st-triangstep}

Starting with a geometric object  ${\cal O}= \{{\cal O}_1 = (S_1, I_1, f_1), {\cal O}_2 = (S_2,
I_2, f_2), \ab \ldots, \ab {\cal O}_n = \ab (S_n, I_n, f_n)\}$, the partitioning algorithm
identifies a list $\chi$ of time moments that is used to partition the time domain $I =
\overline{\bigcup}_{i=1}^nI_i$ of ${\cal O}$ into points and open intervals. For each element in
that partition (point or open interval), we now triangulate the part of ${\cal O}$ restricted to
that point or open interval.

The triangulation of the snapshots of ${\cal O}$ at the time moments in $\chi$ is straightforward.
For each of the time moments $\tau$ of $\chi$, the spatial triangulation method ${\cal T}_S$ is
applied to the snapshot ${\cal O}^\tau$. For each of the triangles $\trc{}$ in ${\cal T}_S({\cal
O}^\tau)$, an atomic object is constructed with $\trc{}$ as reference object, the singleton
$\{\tau\}$ as time domain and the identity as its transformation function.

The triangulation of the parts of ${\cal O}$ restricted to the open intervals in the time partition
requires a new technique. We can however benefit from the fact that throughout each interval, all
snapshots of ${\cal O}$ have an ${\cal T}_S$-isomorphic triangulation. For each of the open
intervals defined by two subsequent elements $]\tau_j, \tau_{(j+1)}[$ of $\chi$, we compute the
snapshot at the middle $\tau_m = \frac{1}{2}(\tau_j + \tau_{(j+1)})$ of $]\tau_j, \tau_{(j+1)}[$
and its triangulation ${\cal T}_S({\cal O}^{\tau_m})$. Each triangle boundary segment that
contributes to the boundary of ${\cal O}^{\tau_m}$ at time moment $\tau_m$, will also contribute to
the boundary of ${\cal O}$ at the snapshot of ${\cal O}$ at any time moment $\tau \in ]\tau_j,
\tau_{(j+1)}[$.  So, the moving line segment can be considered a boundary segment throughout
$]\tau_j, \tau_{(j+1)}[$. If two carriers of boundary segments intersect at time moment $\tau_m$,
the intersection of the moving segments will exist throughout $]\tau_j, \tau_{(j+1)}[$, and so on.
Therefor, we will compute the spatial triangulation of the snapshot ${\cal O}^{\tau_m}$ using the
procedure ${\cal T}_S$, but we will  copy every action on a point or line segment at time moment
$\tau_m$ on the moving point of line segment of which the point or segment is a snapshot. The
triangles returned by the spatial triangulation algorithm when applied to ${\cal O}^{\tau_m}$ will
be reference objects for the atomic objects, returned by the \st triangulation algorithm. These
atomic objects exist during the interval $]\tau_j, \tau_{(j+1)}[$. Knowing the functions
representing the time-dependent corner points of the triangles (because of the copying), together
with the time interval and the reference object, we can deduce the transformation function and
construct atomic objects (the formula computing this transformation was given in
\cite{amai03}).

Next, a detailed description of the \st triangulation is given in Algorithm~\ref{algo-triang-st}.
In this description of the \st triangulation procedure, we will use the data type \emph{Points}
which is a structure containing a (\ndim{2}) point (represented using a pair of real numbers), a
pair of rational functions of $t$ (a rational function is represented using a pair of vectors of
integers, denoting the coefficients of a polynomial), representing a moving point, and finally a
time interval (represented as a pair of real numbers and two flags indicating whether the interval
is open or closed at each end point). We will only use or fill in this time information when
mentioned explicitly. Given an element $Pt$ of type \emph{Points}, we address the point it stores
by $Pt\rightarrow Point$, the functions of time by $Pt\rightarrow f_x$ and $Pt\rightarrow f_y$
respectively, and the begin and end point of the time interval by $Pt\rightarrow I_b$ and
$Pt\rightarrow I_e$. The flags $Pt\rightarrow C_b$ and $Pt\rightarrow C_e$ are true when the
interval is closed at its begin or end point respectively. A pair of elements of the type
\emph{Points} is denoted an element of the type \emph{Segments}.

\begin{algorithm}\rm
\caption{\textbf{Triangulate} (Input: ${\cal O}=\ab\{{\cal O}_1,\ab...,\ab{\cal O}_n\}, \chi =
(\tau_1, \ldots, \tau_k)$, Output = $\{{\cal O}'_1, \ldots, {\cal O}'_\ell\}$)}
\label{algo-triang-st}
\begin{algorithmic}[1]
\FORALL{time moments $\tau_j, j = 1 \ldots k$, of $\chi$}
    \FORALL{triangles $T$ in ${\cal T}_S({\cal O}^{\tau_j})$}
        \STATE return the atomic element $(T, \{\tau_j\}, Id)$.
    \ENDFOR
\ENDFOR

\STATE Let $S_<$ be the list containing all atomic objects ${\cal O}_i= (S_i, I_i, f_i), 1 \leq i
\leq n$, sorted by the begin points $I_{i,b}$ of their time domains.

\STATE Let $S_\textrm{Active}$ be a list of elements of the type
 \emph{Segments}, $S_\textrm{Active} = ()$.

\FORALL{pairs $(\tau_j, \tau_{j+1})$, $j= 1 \ldots (k-1))$, in $\chi$}
    \STATE $\tau_m$:= $\frac{1}{2}(\tau_j + \tau_{j+1})$.
    \STATE Remove all elements $(Pt_1, Pt_2)$ of $S_\textrm{Active}$ for which
    $\tau_j = Pt_1\rightarrow I_e = Pt_2\rightarrow I_e$.
    \FORALL{elements $(Pt_1, Pt_2)$ remaining in $S_\textrm{Active}$}
        \STATE $Pt_r\rightarrow Point$ := $(Pt_r\rightarrow f_x(\tau_m), Pt_r\rightarrow
        f_y(\tau_m))$, $r = 1,2$.
    \ENDFOR
    
    \FORALL{${\cal O}_i = (S_i = (\vect{a}_1, \vect{a}_2, \vect{a}_3), I_i, f_i)$ in $S_<$ for which $I_{i,b}$ is $\tau_i$ }
        \STATE Construct three \emph{Points} $Pt_1$, $Pt_2$ and $Pt_3$ such that $Pt_r\rightarrow Point =
        \vect{a}_r$, $Pt_r\rightarrow f_x = f_i(a_{r,x}, \tau_m)$, $Pt_r\rightarrow f_y = f_i(a_{r,y},
        \tau_m)$ and $Pt_r\rightarrow I_b$ and $Pt_r\rightarrow I_e$ respectively contain $\tau_j$
        and $\tau_{j+1}$ ($r = 1, \ldots, 3$).
        \STATE Construct three \emph{Segments} $St_1$, $St_2$ and $St_3$, containing two different
        elements from the set $\{Pt_1, Pt_2, Pt_3\}$. Add them to $S_\textrm{Active}$.
    \ENDFOR
    \STATE Compute the set ${\cal B}^t(S_\textrm{Active})$ of elements of the type \emph{Segments}, using only the constant point information of the elements of $S_\textrm{Active}$.
    Meanwhile, construct the subdivision ${\cal U}({\cal O}^{\tau_m})$.
    \STATE Compute the convex hull ${\cal CH}^t(S_\textrm{Active})$, using only the constant point information of the elements of $S_\textrm{Active}$, a list of elements of the type \emph{Points}.
    \STATE Construct \mbox{DCEL}$^t(S_\textrm{Active})$, where each half-edge (resp. origin) is now an
    element of the type \emph{Segments} (resp. \emph{Points}). Use ${\cal CH}^t(S_\textrm{Active})$ as a bounding box. Each time the
    intersection of two constant carriers is computed, also compute the formula representing the
    moving intersection point.
    \WHILE {there are any unvisited \emph{Segments} $St$ in \mbox{DCEL}$^t({\cal S})$ left}
    \STATE Compute the list E$^t$list of \emph{Segments} that form a convex polygon. Compute the
    \emph{Points} structure $Pt_m$ containing both the constant and time-dependent center of mass of
    that polygon
    \IF {$Pt_m\rightarrow Point$  belongs to a face of $U({\cal O}^{\tau_m})$}
        \FORALL {elements $St = (Pt_1, Pt_2)$ of E$^t$list}
       \STATE  Output the atomic object $(S, I, f)$, where $S$ is the triangle with corner points
       $Pt_1\rightarrow Point$, $Pt_2\rightarrow Point$ and $Pt_m\rightarrow Point$ and $I$ is $]\tau_j,
       \tau_{j+1}[$. The transformation function $f$ is computed using the functions $Pt_1\rightarrow
       f_x$, $Pt_1\rightarrow f_y$, $Pt_2\rightarrow f_x$, $Pt_2\rightarrow f_y$, $Pt_m\rightarrow f_x$
       and $Pt_m\rightarrow f_y$.
        \ENDFOR
    \ENDIF
\ENDWHILE

\ENDFOR
\end{algorithmic}
\end{algorithm}

We again illustrate the \st triangulation algorithm on the geometric object of
example~\ref{ex-running-st}.

\begin{figure}
\centerline{\includegraphics[width=350pt, height = 150pt]{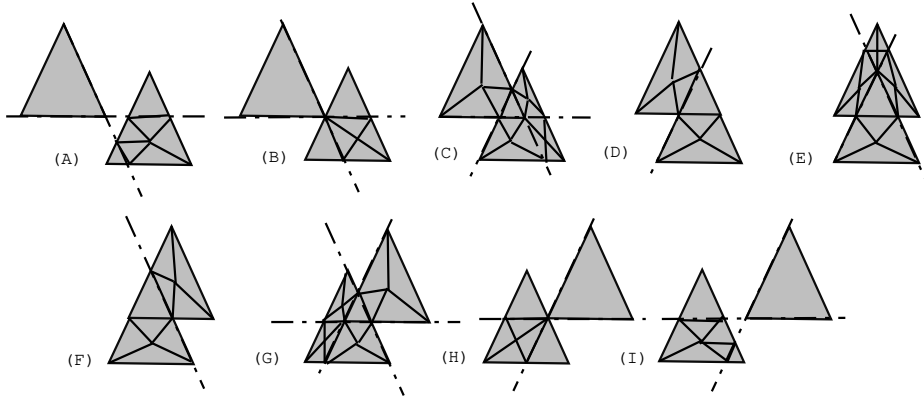}}
  \caption{The triangulations of the objects of Example~\ref{ex-running-st} at time moments $t = \frac{1}{4}$ {\tt(A)}, $t = \frac{1}{2}$ {\tt(B)}, $t = 1$ {\tt(C)}, $t = \frac{3}{2}$ {\tt(D)}, $t = 2$ {\tt(E)}, $t = \frac{5}{2}$ {\tt(F)}, $t = 3$ {\tt(G)}, $t = \frac{7}{2}$ {\tt(H)} and $t = 4$ {\tt(I)}.}\label{fig-triang-st}
\end{figure}

\begin{example}\rm\rm\rm\label{ex-running-triang}
Recall from Example~\ref{ex-running-st} that ${\cal O} = \{{\cal O}_1, {\cal O}_2\}$, where ${\cal
O}_1$ is given as $ (((-1,\ab 0),\ab (1,\ab 0),(0,2)),\ab[0,4], \ab Id)$ and ${\cal O}_2$ is given
as $ (((-3,\ab 1),\ab (-1,\ab 1),\ab (-2,\ab 3)),\ab [0,4], f)$ and $f$ is the affinity 
mapping triples $(x, y, t)$ to pairs $(x+t, y)$.

From Example~\ref{ex-running-partition}, we recall that the output of the procedure {\bf Partition}
on input ${\cal O}$ was the list $\chi = (0, \frac{1}{2}, \frac{3}{2}, \frac{5}{2}, \frac{7}{2},
4)$.

The triangulation of the snapshots at one of the time moments in $\chi$ are shown in
Figure~\ref{fig-triang-st}. To keep the example as simple as possible, we did not further
triangulate convex polygons that are triangles already.

The open intervals to be considered are $]0,\frac{1}{2}[$, $]\frac{1}{2},\frac{3}{2}[$,
$]\frac{3}{2},\frac{5}{2}[$, $]\frac{5}{2},\frac{7}{2}[$ and $]\frac{7}{2},4[$. We illustrate the
triangulation of the interval $]0,\frac{1}{2}[$. During the time interval $]0,\frac{1}{2}[$, the
triangulation will always look like the one shown in Part (A) of Figure~\ref{fig-triang-st}. Hence,
${\cal O}_2$ will not change, and ${\cal O}_1$ will be partitioned into seven triangles. The top
one will not change, so the atomic object $(((0,2),(1, \frac{-1}{2}),(1, \frac{1}{2})),
]0,\frac{1}{2}[, Id)$ will be part of the output. For the others, we have to compute the
time-dependent intersections between the carriers and afterwards apply the formula from
\cite{amai03}. We illustrate this for ${\cal O}_2$. the snapshot of ${\cal O}_2$ at the middle
point $\frac{1}{4}$ of $]0, \frac{1}{2}[$ is the triangle with corner points $(\frac{-11}{4}, 1)$,
$(\frac{-3}{4}, 1)$ and $(\frac{-7}{4}, 3)$. Its time-dependent corner points are $(-3+t,1)$,
$(-1+t, 1)$ and $(-2+t,3)$. Solving the matrix equation

$$ \left(
\begin{array}{@{}cccccc@{}}
\frac{-11}{4} & 1 & 0 & 0 & 1 &0\\ 0 &0 & \frac{-11}{4} & 1 & 0 & 1\\ \frac{-7}{4} & 3 & 0 & 0 & 1 &0\\
0 &0 & \frac{-7}{4} & 3 & 0 & 1\\ \frac{-3}{4} & 1 & 0 & 0 & 1 &0\\ 0 &0 & \frac{-3}{4} & 1 & 0 &
1\
 \end{array}
 \right) \left(
\begin{array}{@{}c@{}}
a(t)\\  b(t) \\ c(t)\\  d(t)\\ e(t) \\ f(t) \\
 \end{array}
 \right) = \left(
\begin{array}{@{}c@{}}
-3+t\\  1 \\
-2+t\\  3\\
-1+t \\ 1\\
 \end{array}
 \right)$$

\par\noindent gives the transformation function $f'$ that maps triples $(x, y, t)$ to pairs $(x -\frac{1}{4} +
t, y)$. \qed\end{example}

We also give the output complexity and time complexity for this triangulation step.

\begin{property}[\textbf{Triangulation step: output complexity}]\rm
\label{prop-complex-output-st-triang} Given a geometric object ${\cal O}= \{{\cal O}_1 = (S_1, I_1,
f_1), \ab{\cal O}_2 = \ab(S_2, I_2, f_2), \ab\ldots, \ab {\cal O}_n = \ab (S_n, \ab I_n, \ab
f_n)\}$ consisting of $n$ atomic objects and a finite partition $\chi$ of its time domain into $k$
time points and $k - 1$ open intervals. The procedure {\bf Triangulation}, as described in
Algorithm~\ref{algo-triang-st}, returns $O(n^2k)$ atomic objects. \end{property}
\par\noindent{\bf Proof.} The number of atomic objects returned by the triangulation
procedure for one time interval is the same as the number of triangles returned by the spatial
triangulation method on a snapshot in that interval. We know from
Property~\ref{prop-complex-output--triang-snapshot} that the number of triangles in the
triangulation of a snapshot composed from $n$ triangles is $O(n^2)$. Since their are $O(k)$ moments
and intervals for which we have to consider such a triangulation, or a slightly adapted version of
it, this gives $O(n^2k)$.\qed

\medskip

\begin{property}[\textbf{Triangulation step: computational complexity}]\rm
\label{prop-complex-time-st-triang} Let  
a geometric object ${\cal O}= \{{\cal O}_1 = (S_1, I_1,
f_1), {\cal O}_2 = (S_2, I_2, f_2), \ldots, \ab {\cal O}_n = \ab (S_n, \ab I_n, \ab f_n)\}$ consisting of
$n$ atomic objects and a finite partition $\chi$ of its time domain into $k$ time points and $k -
1$ open intervals be given. 
Let $d$ be the maximal degree of any polynomial in the definition of the
transformation functions $f_i, 1 \leq i \leq n$ and let $\epsilon$ be the desired precision for
computing the zeros of polynomials.  The procedure {\bf Partition}, as described in
Algorithm~\ref{algo-triang-st-split}, returns a \st triangulation of ${\cal O}$ in time $O(kz(d,
\epsilon)n^2\log n)$. \end{property}
\par\noindent{\bf Proof.} The first for-loop of Algorithm~\ref{algo-triang-st} is executed $k$ times.
The time needed for computing the snapshot of one atomic object at a certain time moment is $z(d,
\epsilon)$. The spatial triangulation algorithm ${\cal T}_S$ runs in time $O(n^2\log n)$, as was
shown in Property~\ref{prop-complex-time-triang-snapshot}. So we can conclude that the body of the
first for-loop needs $O(n^2\log n + nz(d, \epsilon))$ time. Sorting the atomic objects by their
time domains takes $O(n\log n)$.

The second for loop is executed once for each open interval, defined by two consecutive elements of
$\chi$. In the body of this loop, first the list $S_\textrm{Active}$ is updated. Each insertion or
update takes time $z(d, \epsilon)$. At most all objects are in the list $S_\textrm{Active}$, so
this part, described in the Lines 10 through 18 of Algorithm~\ref{algo-triang-st}, needs time
$O(nz(d, \epsilon))$. The next part, described in the Lines 19 through 29 essentially is the
spatial triangulation algorithm, but, any time the intersection between two line segments is
computed, also the rational functions defining the time-dependent intersection of their associated
time-dependent line segments are computed. Computing those functions takes time $z(d, \epsilon)$.
So the second part of the body of the second for loop requires $O(z(d, \epsilon)n^2\log n)$.

If we add up the time complexity of two for-loops and the sorting step, we have $O(k(nz(d,
\epsilon) + n^2\log n) + n\log n + kz(d, \epsilon)(n + n^2\log n))$, which is $O(kz(d,
\epsilon)n^2\log n)$. \qed

\subsection{The Merge Step.}\label{sec-triang-st-mergestep}

We already mentioned briefly in the description of the partitioning step that the partition of the
time domain, as computed by Algorithm~\ref{algo-triang-st-split}, might be finer than necessary.
The partitioning algorithm takes into account all line segments, also those of objects that, during
some time span, are entirely overlapped by other objects. \ignore{We want our triangulation to be
affine-invariant which means, amongst other things, that it should be independent of the
representation of a \st object by means of atomic objects. So whether some unnecessary object is
present or not, the partition of the time domain has to be the same.} To solve this, we merge as
much elements of the time partition as possible.

The partition of the time domain is such that the merging algorithm will either try to merge a time
point $\tau$ and an interval of the type $]\tau, \tau')$ or $(\tau', \tau[$, or two different
intervals of the type $(\tau'',\tau]$ and $]\tau, \tau')$ or $(\tau'', \tau[$ and $[\tau, \tau')$.
Here, we use the (unusual) 
notational convention that $($ and $)$ can be either $[$ or $]$.

The simplest case is when a time moment and an interval have to be tested. Assume that these are
$(\tau', \tau[$ and $\tau$, respectively. These elements can be merged if there is a one to one
mapping $M$ from the atomic elements with time domain $(\tau', \tau[$ to those with time domain
$\{\tau\}$ in the triangulation. Furthermore, for each pair of atomic objects ${\cal O}_1 = (S_1,
(\tau', \tau[, f_1)$ and ${\cal O}_2 = (S_2, \{\tau\}, Id)$, ${\cal O}_2 = M({\cal O}_1)$ if and
only if the left limit $\lim_{t\to\tau}f_1(S_1, t) = S_2$. Note that, for rational functions $f$ of
$t$, $\lim_{t\to\tau}f(t)$ equals $f(\tau)$, provided that $\tau$ is in the domain of
$f$~\footnote{Note that $\tau$ is in the domain of $f$ only if all coefficients of the
transformation function $f$ are well-defined for $t = \tau$ and if the determinant of $f$ is
nonzero for $t = \tau$.}.

When two intervals are to be merged, the procedure involves some more tests. Let $(\tau'', \tau[$
and $[\tau, \tau')$ be the intervals to be tested. First, we have to verify that for each atomic
object ${\cal O}_1 = (S_1, (\tau'', \tau[ , f_1)$, $[\tau, \tau')$ is in the domain of $f_1$ and
that for each atomic object ${\cal O}_2 = (S_2, [\tau, \tau') , f_2)$, $(\tau'', \tau[$ is in the
domain of $f_2$. Second, we have to test whether $(\tau'', \tau[$ can be continuously expanded to
$(\tau'',\tau]$. This involves the same tests as for the simple case where an interval and a point
are tested. Finally, two atomic object can only be merged if the combined atomic object again is an atomic object. This means that, if $S_2$ would have been
chosen as a reference object for ${\cal O}_1$, then $f_1$ would be equal to $f_2$, and vice versa.
This can be tested (\cite{amai03}).

This merge step guarantees that the atomic objects exist maximally and that the resulting
triangulation is the same for geometric objects that represent the same \st object.
Algorithm~\ref{algo-triang-st-merge} shows this merging step in detail.

\begin{algorithm}\rm
\caption{\textbf{Merge} (Input: ${\cal O}=\ab\{{\cal O}_1,\ab...,\ab{\cal O}_n\}, \chi = (\tau_1,
\tau_2, \ldots, \tau_k)$, Output: $\{{\cal O}'_1, {\cal O}'_2, \ldots, {\cal O}'_\ell\}$}
\label{algo-triang-st-merge}
\begin{algorithmic}[1]
\STATE Sort all atomic objects ${\cal O}_i$ by their time domains.

\STATE Let $\chi'$ be the list $(\tau_1, ]\tau_1, \tau_2[, \tau_2, \ldots, ]\tau_{k - 1}, \tau_k[,
\tau_k)$.

\STATE Let $J_1$ be the first element of $\chi'$ and $J_2$ the second.

\WHILE{there are any elements in $\chi'$ left}
    \STATE ${\cal S}_1$  (resp. ${\cal S}_2$) is the set of all objects having $J_1$ (resp. $J_2$) as their time
    domain.
    \IF{$J_1$ is a point}
        \STATE Preprocess the reference objects of the elements of ${\cal S}_1$ such that we can
        search the planar subdivision ${\cal U}_1$  they define.
        \STATE let {\bf Found} be true.
        \FORALL{objects ${\cal O}_i = (S_i, J_2, f_i)$ in ${\cal S}_2$}
            \STATE Check whether $J_1$ is part of the time domain of $f_i$.
            \STATE Compute their snapshot at time $J_1$ (which is a triangle $\trc{}$).
            \STATE Do a point location query with the center of mass of
            $\trc{}$ in ${\cal U}_1$ and check
            whether the triangle found in ${\cal S}_1$ has the same
            coordinates as $\trc{}$. If not, {\bf Found} becomes
            false.
            \IF{ {\bf Found} is false}
                \STATE break;
            \ENDIF
        \ENDFOR
        \IF{ {\bf found} is true}
            \STATE remove all elements of ${\cal S}_1$ from ${\cal O}$ and extend the time domain of all elements of ${\cal S}_2$
            to $J_1 \cup J_2$.
            \STATE $J_1 = J_1 \cup J_2$ and $J_2$ is the next element of
            $\chi'$ if any exists.
        \ELSE
            \STATE $J_1 = J_2$ and $J_2$ is the next element of
            $\chi'$, if any exists.
        \ENDIF
    \ELSE
        \IF{$J_2$ is a point}
            \STATE do the same as in the previous case, but switch
            the roles of $J_1$ and $J_2$.
        \ELSE
            \STATE Let $J'_1$ be the element of $\{J_1, J_2\}$ the form $(\tau'', \tau[$
and $J'_2$ the one of the form $[\tau, \tau')$.
            \STATE Check whether $(\tau'', \tau[$ and $\{\tau\}$ can be merged.
            \IF{this can be done}
            \STATE Check for each pair of matching atomic objects whether their transformation
            functions are the same (\cite{amai03}).
            \ENDIF
        \ENDIF
    \ENDIF
\ENDWHILE
\end{algorithmic}
\end{algorithm}

We illustrate Algorithm~\ref{algo-triang-st-merge} on the geometric object of
Example~\ref{ex-running-st}.

\begin{example}\rm\rm\rm\label{ex-running-merge}
Recall from Example~\ref{ex-running-st} that ${\cal O} = \{{\cal O}_1, {\cal O}_2\}$, where ${\cal
O}_1$ is given as $ (((-1,\ab 0),\ab (1,\ab 0),(0,2)),\ab[0,4], \ab Id)$ and ${\cal O}_2$ is given
as $ (((-3,\ab 1),\ab (-1,\ab 1),\ab (-2,\ab 3)),\ab [0,4], f)$ and $f$ is the 
linear affinity
mapping triples $(x, y, t)$ to pairs $(x+t, y)$.

From Example~\ref{ex-running-partition}, we recall that the output of the procedure {\bf partition}
on input ${\cal O}$ was the list $\chi = (0, \frac{1}{2}, \frac{3}{2}, \frac{5}{2}, \frac{7}{2},
4)$. This resulted in a partition of the interval $[0,4]$ consisting of the elements $\{0\}$,
$]0,\frac{1}{2}[$, $\{\frac{1}{2}\}$, $]\frac{1}{2},\frac{3}{2}[$, $\{\frac{3}{2}\}$,
$]\frac{3}{2},\frac{5}{2}[$, $\{\frac{5}{2}\}$, $]\frac{5}{2},\frac{7}{2}[$, $\{\frac{7}{2}\}$,
$]\frac{7}{2},4[$ and $\{4\}$. For each of these elements, (a snapshot of) their triangulation is
shown in Figure~\ref{fig-triang-st}.

During the merge step, the elements $t = 0$ and $]0, \frac{1}{2}[$ of the time partition will be
merged. \qed\end{example}

It is straightforward that the output and input of the merging algorithm have the same order of
magnitude. Indeed, it is possible that no intervals are merged, and hence no objects. We discuss
the computational complexity of the algorithm next. Note that the complexity is expressed in terms
of the size of the input to the merging algorithm, which is the output of the \st triangulation
step.

\begin{property}[\textbf{Merge step: computational complexity}]\rm
\label{prop-complex-time-merge} Given a geometric object ${\cal O}= \{{\cal O}_1 = (S_1, I_1, f_1),
{\cal O}_2 = (S_2, I_2, f_2), \ldots, \ab {\cal O}_n = \ab (S_n, I_n, f_n)\}$, which is the output
of the triangulation step, and a finite partition $\chi$ of its time domain into $K$ time moments
and open intervals. Let $d$ be the maximal degree of any polynomial in the definition of the
transformation functions $f_i (1 \leq i \leq n)$ and let $\epsilon$ be the desired precision for
computing the zeros of polynomials. The procedure {\bf Merge}, as described in
Algorithm~\ref{algo-triang-st-merge}, merges the atomic objects in ${\cal O}$ in time $O(n\log
\frac{n}{K} + nz(d, \epsilon))$. \end{property}
\par\noindent{\bf Proof.} Sorting all atomic objects by their time domains can be done in time $O(n\log
n)$. Computing the list $\chi'$ can be straightforwardly done in time $O(K)$. This list will
contain $2K-1$ elements. We assume that $K > 1$ (in case $K=1$ the merging algorithm is not
applied). The while-loop starting at Line 4 of Algorithm~\ref{algo-triang-st-merge}, is executed at
most $2K-2$ times. Indeed, at each execution of the body of the while-loop, one new element of
$\chi'$ is considered. The {\bf if-else} structure in the body of the while-loop distinguishes
three cases. All cases have the same time complexity, as they are analogous. We explain the first
case in detail.

The number of atomic objects having the same time domain is of the order of magnitude of
$O(\frac{n}{K})$. This follows from Property~\ref{prop-complex-output-st-triang}.
 The preprocessing of the snapshot takes $O(\frac{n}{K})$ time~\cite{pointlocation}. The for-loop, starting at Line 9 of Algorithm~\ref{algo-triang-st-merge}
is executed at most $O(\frac{n}{K})$ times. The time needed for checking whether an atomic object
exists at some time moment and computing the snapshot (a triangle) is $z(d, \epsilon)$. Because of
the preprocessing on the snapshot at time moment $J_1$, testing the barycenter of the triangle
against that snapshot can be done in $O(\log \frac{n}{K})$ time~\cite{pointlocation}. In case the
snapshots are the same, adjusting the time domains of all atomic objects takes time
$O(\frac{n}{K})$. Summarizing, the time complexity of the first case is $O(\frac{n}{K}\log
\frac{n}{K} + \frac{n}{K}z(d, \epsilon))$.

Combining this with the fact that the while-loop is executed $O(K)$ times, and the time complexity
of the first two steps of the algorithm, we get an overall time complexity of $O(n\log \frac{n}{K}
+ nz(d, \epsilon))$.\qed

\medskip

Finally, the \st triangulation procedure ${\cal T}_{ST}$ combines the partition, triangulation and
merging step. Algorithm~\ref{algo-triang-st-all} combines all steps.

\begin{algorithm}\rm
\caption{${\cal T}_{ST}$ (Input: ${\cal O}=\ab\{{\cal O}_1,\ab...,\ab{\cal O}_n\}$, Output:
$\{{\cal O}'_1, {\cal O}'_2, \ldots, {\cal O}'_\ell\}$)} \label{algo-triang-st-all}
\begin{algorithmic}[1]
\STATE $\chi$ = {\bf Partition}({\cal O});

\STATE $\{{\cal O}''_1, {\cal O}''_2, \ldots, {\cal O}''_m\}$ = {\bf Triangulate}(${\cal O}$,
$\chi$);

\IF{$\chi$ has more than one element}
    \STATE $\{{\cal O}'_1, {\cal O}'_2, \ldots, {\cal O}'_\ell\}$ = {\bf
Merge}($\{{\cal O}''_1, {\cal O}''_2, \ldots, {\cal O}''_m\}$, $\chi$);
    \STATE return $\{{\cal O}'_1, {\cal O}'_2, \ldots, {\cal O}'_\ell\}$.
\ELSE
    \STATE return $\{{\cal O}''_1, {\cal O}''_2, \ldots, {\cal O}''_m\}$.
\ENDIF
\end{algorithmic}
\end{algorithm}

\medskip

The following property follows from Property~\ref{prop-complex-output-partition} and
Property~\ref{prop-complex-output-st-triang}.

\begin{property}[\textbf{${\cal T}_{ST}$:  output complexity}]\rm
\label{prop-complex-output-st-triang-all} Given a geometric object ${\cal O}= \{{\cal O}_1 = (S_1,
I_1, f_1), {\cal O}_2 = (S_2, I_2, f_2), \ldots, \ab {\cal O}_n = \ab (S_n, I_n, f_n)\}$ consisting
of $n$ atomic objects. Let $d$ be the maximal degree of any polynomial in the definition of the
transformation functions $f_i (1 \leq i \leq n)$. The \st triangulation method ${\cal T}_{ST}$, as
described in Algorithm~\ref{algo-triang-st-all}, returns $O(n^5d)$ atomic objects. \end{property}

\medskip
The next property follows from Property~\ref{prop-complex-time-partition},
Property~\ref{prop-complex-time-st-triang} and Property~\ref{prop-complex-time-merge}.
Table~\ref{table-st-all-complex} summarizes the time complexity of the different steps.

\begin{table}
\begin{center}
\begin{tabular}{|l|c|c|}\hline
Step & {Time complexity} & {Output complexity} \\ \hline\hline {\bf Partition} & $O(z(d, \epsilon)
n^3\log n)$ & $O(n^3d)$ \\\hline {\bf Triangulate} &  $O(z(d, \epsilon)dn^5\log n)$ &  $O(n^5d)$
\\\hline {\bf Merge} & $O(n^5d(\log n + z(d, \epsilon)))$ & -\\\hline\hline
Overall  & $O(z(d, \epsilon)dn^5\log n)$ & $O(n^5d)$\\\hline
\end{tabular}
\caption{The output and time complexity of the various parts of Algorithm~\ref{algo-triang-st-all},
when the input is a geometric object, composed of $n$ atomic objects, where the maximal degree of the polynomials
describing the transformation functions is $d$ and the desired precision for computing the zeros of
polynomials is $\epsilon$.}\label{table-st-all-complex}
\end{center}
\end{table}

\begin{property}[\textbf{${\cal T}_{ST}$: computational complexity}]\rm
\label{prop-complex-time-st-triang-all} Given a geometric object ${\cal O}= \ab\{{\cal O}_1 = \ab
(S_1, \ab I_1, \ab f_1), \ab {\cal O}_2 = \ab (S_2, \ab I_2, \ab f_2), \ab\ldots, \ab {\cal O}_n =
\ab (S_n, \ab I_n, \ab f_n)\}$ consisting of $n$ atomic objects. Let $d$ be the maximal degree of
any polynomial in the definition of the transformation functions $f_i (1 \leq i \leq n)$ and let
$\epsilon$ be the desired precision for computing the zeros of polynomials. The \st triangulation
method ${\cal T}_{ST}$, as described in Algorithm~\ref{algo-triang-st-all}, returns a \st
triangulation of ${\cal O}$ in time $O(z(d, \epsilon)dn^5\log n)$. \end{property}

\medskip

We now show that Algorithm~\ref{algo-triang-st-all} describes an affine-invariant \st triangulation
meth\-od.  We remark first that the result of the procedure ${\cal T}_{ST}$ is a \st triangulation.
Given a geometric object ${\cal O}$. It is clear that each snapshot of ${\cal T}_{ST}({\cal O})$ is
a spatial triangulation. Also, $st({\cal O}) = st({\cal T}_{ST}({\cal O}))$. This follows from the
fact that the time partition covers the whole time domain of ${\cal O}$ and that the method ${\cal
T}_S$ produces a spatial triangulation.

 \begin{property}[\textbf{${\cal T}_{ST}$ is affine-invariant}]\rm\label{prop-aff-inv-triang-st}
The \st triangulation meth\-od ${\cal T}_{ST}$, described in Algorithm~\ref{algo-triang-st-all}, is
affine-invariant.
\end{property}
\par\noindent{\bf Proof.} (Recall Definition~\ref{def-affine-invar-triang} for affine-invariance.)
 Let ${\cal O} = \{{\cal O}_1,\ab \ldots, \ab {\cal O}_n\}$ and ${\cal O}' =  \{{\cal O}'_1,\ldots,
{\cal O}'_m\}$ be geometric objects for which for each moment $\tau_0$ in their time domains, there
is an affinity $\alpha_{\tau_0} :\Rn{2}\rightarrow \Rn{2}$ such that 
$\alpha_{\tau_0}(\{{\cal
O}_1,\ldots, {\cal O}_n\}^{\tau_0})=\{{\cal O}'_1,\ldots, {\cal O}'_m\}^{\tau_0}$.

It follows from the construction of the \st triangulation that ${\cal T}_{ST}(\{{\cal O}_1,\ab
\ldots, \ab {\cal O}_n\})^{\tau_0}=$  ${\cal T}_{S}(\{{\cal O}_1,\ab \ldots, \ab {\cal
O}_n\}^{\tau_0})$ and also ${\cal T}_{ST}(\{{\cal O}'_1,\ab \ldots, \ab {\cal O}'_m\})^{\tau_0}=$ 
${\cal T}_{S}(\{{\cal O}'_1,\ab \ldots, \ab {\cal O}'_n\}^{\tau_0})$. The property now follows from
the affine-invariance of the spatial triangulation method ${\cal T}_S$.\qed

\medskip

The following corollary follows straightforwardly from Property~\ref{prop-aff-inv-triang-st}:

\begin{corollary}\rm
Let ${\cal O} = \{{\cal O}_1, \ldots, {\cal O}_n\}$ and ${\cal O}'= \{{\cal O}'_1, \ab \ldots, \ab
{\cal O}'_m\}$ be two geometric objects such that there is an affinity $\alpha: \Rn{2} \rightarrow
\Rn{2}$ such that, for each moment $\tau_0$ in their time domains $\alpha({\cal O}^{\tau_0}) =
{\cal O}'^{\tau_0}$ holds. Then, for each atomic element $(S, I, f)$ of ${\cal T}_{ST}({\cal O})$,
the element $(\alpha(S), I, f)$ belongs to ${\cal T}_{ST}({\cal O}')$.\qed
\end{corollary}

 This shows that the partition is independent of the coordinate system used to represent
the spa\-tio-tem\-po\-ral object. The affine partitions of two spa\-tio-tem\-po\-ral objects that
are affine images of each other only differ in the coordinates of the spatial reference objects of
the atomic objects.

\ignore{
 We now define what is means for a geometric object to be in \emph{${\cal T}_{ST}$-normal
form}.

\begin{definition}\rm\rm
Let ${\cal O}$ be a geometric object given as the set of atomic objects $\{{\cal O}_1, \ldots,
{\cal O}_n\}$. We say that ${\cal O}$ (or more precisely, this set of atomic objects) is in
\emph{${\cal T}_{ST}$ normal form}, if ${\cal T}_{ST}$ on input ${\cal O}$ returns $\{{\cal O}_1,
\ldots, {\cal O}_n\}$ as output.
\end{definition}
}

Remark also that, in practice, either most of the original objects have the same time domain, making the number of intervals in the partition very
small, or all different time domains, which greatly reduces the number of objects existing during each interval. So, in practice, the performance
will be better than the worst case suggests.

\section{Applications}\label{sec-app}
We now describe some applications that we believe can benefit from the triangulation described in
Algorithm~\ref{algo-triang-st-all}. We first say what we mean by a spatio-temporal database.

\begin{definition}\rm\rm
A \emph{spatio-temporal database} is a set of  geometric objects.\qed
\end{definition}

For this section, we assume that each atomic object is labelled with the id of the geometric object
it belongs to.

\subsection{Efficient Rendering of Objects}
When a geometric object that is not in normal form has to be displayed to the user, there are two tasks to
perform. First, the snapshots of the geometric object at each time moment in the time domain of the
object have to be computed. This can be done in a brute force way by computing the snapshots of all
atomic objects. Since some will be empty, this approach might lead to a lot of unnecessary
computations. Another algorithm could keep track of the time domains of the individual atomic
objects and keep a list of \emph{active} ones at the moment under consideration, which has to be
updated every instant. If the geometric object is in normal form, the atomic objects can be sorted
by their time domains, and during each interval in the partition of time domain, the list of active
atomic objects will remain the same.

The second task is the rendering of the snapshot. If the geometric object is not in normal form,
the snapshots of the atomic objects overlap, so pixels will be computed more than once. Another
solution is computing the boundary, but this might take too long in real-time applications.
When a geometric object is in normal form, no triangles overlap, so each pixel will be
computed only once.

\subsection{Moving Object Retrieval}
The triangulation provides a means of automatic affine invariant feature extraction for moving
object recognition. Indeed, the number of intervals in the time domain indicates the complexity of
the movement of the geometric object. This can be used as a first criterium for object matching.
For objects having approximately the same number of intervals in their time domains, the snapshots
at the middle of each time interval can be compared. If they are all similar, which can be, for
example, defined as $\cali{T}_S$-isomorphic, the objects match. Or, if more exact comparison is
needed, one can extract an affine-invariant description from the structure of the elements of the
triangulation of the snapshot (see also Section~\ref{queries}).

\subsection{Surveillance Systems}
In so\-me applications, e.g., surveillance systems, it is important to know the time moments when
something changed, when some discontinuity appeared. This could mean that an unauthorized person
entered a restricted area, for example, or that a river has burst its banks. Triangulating the
contours of the recorded images and reporting all single points and end points of intervals of the
partition of the time domain indicates all moments when some discontinuity might have occurred.

\subsection{Precomputing Queries}
\begin{figure}[h]
  \centerline{\includegraphics[width=150pt]{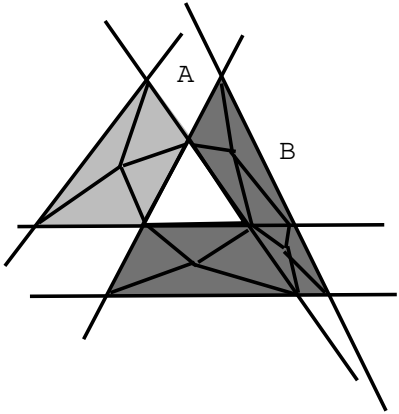}}
  \caption{The set operations between objects are pre-computed in the triangulation.}
  \label{figsetops}
\end{figure}

If we do not triangulate each geometric object in a database separately, but use the contours of
all geometric objects together in the triangulation, the atomic objects in the result will have the
following nice property. For each geometric object in the original database, an atomic object will
either belong to (or be a subset of) it entirely, or not at all. This means that we can label each
element of the spatio-temporal triangulation of the database with the set of id's of the geometric
objects it belongs to. We illustrate this for the spatial case only in Figure~\ref{figsetops}.
Suppose we have two triangles $A$ and $B$. The set $A$ is the union of the light grey and white
parts of the figure, the set $B$ is the union of the dark grey and white parts of the figure. After
triangulation, we can label the light grey triangles with $\{A\}$, the white triangle with $\{A,
B\}$, and the dark grey triangles with $\{B\}$.

Using this triangulation of databases in a preprocessing stage, means that the results of queries
that ask for set operations between geometric objects are also pre-computed. Answering such a query
boils down to checking labels of atomic objects. This means a lot of gain in speed at query time.
Indeed, even to compute, for example, the intersection of two atomic objects, one has to compute
first the intervals where the intersection exists, and then to consider all possible shapes the
intersection can have, and again represent it by moving triangles.

\subsection{Maintaining the Triangulation.} If a geometric object has to be inserted into or removed from a database (i.e., a collection of geometric objects), the triangulation has to be
recomputed for the intervals in the partition of the time domain that contain the time domain of
the object under consideration. This may require that the triangulation in total has to be
recomputed.

However, the nature of a lot of spa\-tio-tem\-po\-ral applications is such that updates involve
only the insertion of objects that exists {\em later} in time than the already present data. In
that case, only the triangulation at the latest time interval of the partition should be recomputed
together with the new object, to check whether the new data are a continuation of the previous.
Also, data is removed only when it is outdated. In that case a whole time interval of data can be
removed. Examples of such spa\-tio-tem\-po\-ral applications are surveillance, traffic monitoring
and cadastral information systems.

\subsection{Affine-invariant Querying of Spatio-\-tem\-po\-ral \\
Databases}~\label{queries}

An interesting topic for further work is to compute a new, affine-invariant description of
geometric objects in normal form, that does not involve coordinates of reference objects. The
structure of the atomic objects in the spatio-temporal triangulation can be used for that. Once
such a description is developed, a query language can be designed that asks for affine-invariant
properties of objects only.

\section{Concluding Remarks}\label{sec-conclusion}
We adopted the hierarchical data model of Chomicki and Revesz~\cite{cr-99} for moving objects, since
it is natural and flexible. However, this model lacks a normal form, as different sets of objects
might represent the same spatio-temporal set in $\R^{2}\times \R$. Furthermore, we are interested in
affine-invariant representation and querying of objects, as the choice of origin and unit of
measure should not affect queries on spatio-temporal data.

We first introduced a new affine-invariant triangulation method for spatial data. We then extended
this method for spa\-tio-tem\-po\-ral data in such a way that the time domain is partitioned in
intervals for which the triangulations of all snapshots are {\em isomorphic}.

The proposed affine-in\-va\-riant triangulation is natural and can serve as an affine-invariant
normal form for spa\-tio-tem\-po\-ral data. Further work includes the affine-invariant
finite representation of data and the design of an affine-generic spatial/ spa\-tio-tem\-po\-ral query
language to query such a normal form.

\bibliographystyle{plain}

\end{document}

%% file: spatialalgo.pstex_t
\begin{picture}(0,0)%
\includegraphics{spatialalgo.pstex}%
\end{picture}%
\setlength{\unitlength}{3947sp}%
\begingroup\makeatletter\ifx\SetFigFont\undefined%
\gdef\SetFigFont#1#2#3#4#5{%
  \reset@font\fontsize{#1}{#2pt}%
  \fontfamily{#3}\fontseries{#4}\fontshape{#5}%
  \selectfont}%
\fi\endgroup%
\begin{picture}(7116,5915)(1249,-7846)
\put(6366,-3013){\makebox(0,0)[lb]{\smash{{\SetFigFont{11}{13.2}{\rmdefault}{\mddefault}{\updefault}{$v_5$}%
}}}}
\put(4481,-7659){\makebox(0,0)[lb]{\smash{{\SetFigFont{11}{13.2}{\rmdefault}{\mddefault}{\updefault}{$v_4$}%
}}}}
\put(4548,-5100){\makebox(0,0)[lb]{\smash{{\SetFigFont{11}{13.2}{\rmdefault}{\mddefault}{\updefault}{$v_2$}%
}}}}
\put(1518,-4966){\makebox(0,0)[lb]{\smash{{\SetFigFont{11}{13.2}{\rmdefault}{\mddefault}{\updefault}{$v_1$}%
}}}}
\put(1384,-7592){\makebox(0,0)[lb]{\smash{{\SetFigFont{11}{13.2}{\rmdefault}{\mddefault}{\updefault}{$v_3$}%
}}}}
\put(2798,-6716){\makebox(0,0)[lb]{\smash{{\SetFigFont{11}{13.2}{\rmdefault}{\mddefault}{\updefault}{$v_5$}%
}}}}
\put(1922,-6380){\makebox(0,0)[lb]{\smash{{\SetFigFont{11}{13.2}{\rmdefault}{\mddefault}{\updefault}{$f_2$}%
}}}}
\put(2865,-5908){\makebox(0,0)[lb]{\smash{{\SetFigFont{11}{13.2}{\rmdefault}{\mddefault}{\updefault}{$f_1$}%
}}}}
\put(3875,-6312){\makebox(0,0)[lb]{\smash{{\SetFigFont{11}{13.2}{\rmdefault}{\mddefault}{\updefault}{$f_3$}%
}}}}
\put(2865,-7053){\makebox(0,0)[lb]{\smash{{\SetFigFont{11}{13.2}{\rmdefault}{\mddefault}{\updefault}{$f_4$}%
}}}}
\put(2730,-5437){\makebox(0,0)[lb]{\smash{{\SetFigFont{11}{13.2}{\rmdefault}{\mddefault}{\updefault}{$e_{2,1}$}%
}}}}
\put(2326,-5639){\makebox(0,0)[lb]{\smash{{\SetFigFont{11}{13.2}{\rmdefault}{\mddefault}{\updefault}{$e_{1,5}$}%
}}}}
\put(3471,-5639){\makebox(0,0)[lb]{\smash{{\SetFigFont{11}{13.2}{\rmdefault}{\mddefault}{\updefault}{$e_{5,2}$}%
}}}}
\put(2730,-4764){\makebox(0,0)[lb]{\smash{{\SetFigFont{11}{13.2}{\rmdefault}{\mddefault}{\updefault}{$(A)$}%
}}}}
\put(1518,-4495){\makebox(0,0)[lb]{\smash{{\SetFigFont{11}{13.2}{\rmdefault}{\mddefault}{\updefault}{$v_3$}%
}}}}
\put(4346,-4495){\makebox(0,0)[lb]{\smash{{\SetFigFont{11}{13.2}{\rmdefault}{\mddefault}{\updefault}{$v_4$}%
}}}}
\put(3605,-3350){\makebox(0,0)[lb]{\smash{{\SetFigFont{11}{13.2}{\rmdefault}{\mddefault}{\updefault}{$T_2$}%
}}}}
\put(1922,-3619){\makebox(0,0)[lb]{\smash{{\SetFigFont{11}{13.2}{\rmdefault}{\mddefault}{\updefault}{$T_1$}%
}}}}
\put(1249,-2407){\makebox(0,0)[lb]{\smash{{\SetFigFont{11}{13.2}{\rmdefault}{\mddefault}{\updefault}{$v_1$}%
}}}}
\put(4009,-2071){\makebox(0,0)[lb]{\smash{{\SetFigFont{11}{13.2}{\rmdefault}{\mddefault}{\updefault}{$v_2$}%
}}}}
\put(6702,-4764){\makebox(0,0)[lb]{\smash{{\SetFigFont{11}{13.2}{\rmdefault}{\mddefault}{\updefault}{$(B)$}%
}}}}
\put(7847,-2138){\makebox(0,0)[lb]{\smash{{\SetFigFont{11}{13.2}{\rmdefault}{\mddefault}{\updefault}{$v_2$}%
}}}}
\put(4750,-2340){\makebox(0,0)[lb]{\smash{{\SetFigFont{11}{13.2}{\rmdefault}{\mddefault}{\updefault}{$v_1$}%
}}}}
\put(6702,-7794){\makebox(0,0)[lb]{\smash{{\SetFigFont{11}{13.2}{\rmdefault}{\mddefault}{\updefault}{$(D)$}%
}}}}
\put(2865,-7794){\makebox(0,0)[lb]{\smash{{\SetFigFont{11}{13.2}{\rmdefault}{\mddefault}{\updefault}{$(C)$}%
}}}}
\put(7951,-4411){\makebox(0,0)[lb]{\smash{{\SetFigFont{11}{13.2}{\rmdefault}{\mddefault}{\updefault}{$v_4$}%
}}}}
\put(5326,-4636){\makebox(0,0)[lb]{\smash{{\SetFigFont{11}{13.2}{\rmdefault}{\mddefault}{\updefault}{$v_3$}%
}}}}
\end{picture}%

%% file: HK-triangulation.bbl
\begin{thebibliography}{10}

\bibitem{time00}
{\em Proceedings of the 7th International Workshop on Temporal Representation
  and Reasoning}. IEEE Computer Society Press, 2000.

\bibitem{bcr}
J.~Bochnak, M.~Coste, and M.-F. Roy.
\newblock {\em Real Algebraic Geometry}, volume~36 of {\em Ergebenisse der
  Mathematik und ihrer Grenzgebiete. Folge 3.}
\newblock Springer-Verlag, 1998.

\bibitem{STDBM99}
M.~H. B{\"o}hlen, Ch.~S. Jensen, and M.~Scholl, editors.
\newblock {\em Spatio-Temporal Database Management, International Workshop
  {STDBM} 1999}, volume 1678 of {\em Lecture Notes in Computer Science}.
  Springer, 1999.

\bibitem{cz-00}
C.~X. Chen and C.~Zaniolo.
\newblock {SQLST}: A spatio-temporal data model and query language.
\newblock In V.~C.~Storey A.~H.~F.~Laender, S.~W.~Liddle, editor, {\em
  Conceptual Modeling, 19th International Conference on Conceptual Modeling
  (ER'00)}, volume 1920 of {\em Lecture Notes in Computer Science}, pages
  96--111. Springer-Verlag, 2000.

\bibitem{amai03}
J.~Chomicki, S.~Haesevoets, B.~Kuijpers, and P.~Revesz.
\newblock Classes of spatiotemporal objects and their closure properties.
\newblock {\em Annals of Mathematics and Artificial Intelligence},
  (39):431--461, 2003.

\bibitem{cr-99}
J.~Chomicki and P.~Revesz.
\newblock A geometric framework for specifying spatiotemporal objects.
\newblock In {\em Proceedings of the 6th International Workshop on Temporal
  Representation and Reasoning}, pages 41--46. IEEE Computer Society Press,
  1999.

\bibitem{collins}
G.E. Collins.
\newblock Quantifier elimination for real closed fields by cylindrical
  algebraic decomposition.
\newblock In H.~Brakhage, editor, {\em Automata Theory and Formal Languages},
  volume~33 of {\em Lecture Notes in Computer Science}, pages 134--183, Berlin,
  1975. Springer-Verlag.

\bibitem{comp-geom-chap8}
M.~de~Berg, M.~van Kreveld, M.~Overmars, and O.~Schwarzkopf.
\newblock Arrangements and duality.
\newblock In {\em Computational Geometry: Algorithms and Applications},
  chapter~8, pages 165--182. Springer-Verlag, 2000.

\bibitem{comp-geom-chap1}
M.~de~Berg, M.~van Kreveld, M.~Overmars, and O.~Schwarzkopf.
\newblock Computational geometry.
\newblock In {\em Computational Geometry: Algorithms and Applications},
  chapter~1, pages 1--17. Springer-Verlag, 2000.

\bibitem{comp-geom}
M.~de~Berg, M.~van Kreveld, M.~Overmars, and O.~Schwarzkopf.
\newblock {\em Computational Geometry: Algorithms and Applications}.
\newblock Springer-Verlag, 2000.

\bibitem{comp-geom-chap2}
M.~de~Berg, M.~van Kreveld, M.~Overmars, and O.~Schwarzkopf.
\newblock Line segment intersection.
\newblock In {\em Computational Geometry: Algorithms and Applications},
  chapter~2, pages 18--43. Springer-Verlag, 2000.

\bibitem{luc-pods}
F.~Dumortier, M.~Gyssens, L.~Vandeurzen, and D.~Van Gucht.
\newblock On the decidability of semi-linearity for semi-algebraic sets and its
  implications for spatial databases.
\newblock In {\em Proceedings of the 16th ACM Symposium on Principles of
  Database Systems}, pages 68--77. ACM Press, 1997.

\bibitem{pointlocation}
H.~Edelsbrunner, L.~J. Guibas, and J.~Stolfi.
\newblock Optimal point location in a monotone subdivision.
\newblock {\em SIAM J. Comput.}, 15(2):317--340, 1986.

\bibitem{es-97}
M.~Erwig and M.~Schneider.
\newblock Partition and conquer.
\newblock In {\em Proceedings of the 3rd International Conference on Spatial
  Information Theory}, volume 1329 of {\em Lecture Notes in Computer Science},
  pages 389--408. Springer, 1997.

\bibitem{CHOROCHRONOS}
A.~Frank, S.~Grumbach, R.~G{\"u}ting, C.~Jensen, M.~Koubarakis, N.~Lorentzos,
  Y.~Manopoulos, E.~Nardelli, B.~Pernici, H.-J. Schek, M.~Scholl, T.~Sellis,
  B.~Theodoulidis, and P.~Widmayer.
\newblock Chorochronos: A research network for spatiotemporal database systems.
\newblock {\em SIGMOD Record}, 28:12--21, 1999.

\bibitem{ghk-01}
F.~Geerts, S.~Haesevoets, and B.~Kuijpers.
\newblock A theory of spatio-temporal database queries.
\newblock In G.~Ghelli and G.~Grahne, editors, {\em Database Programming
  Languages, 8th International Workshop, {DBPL} 2001}, volume 2397 of {\em
  Lecture Notes in Computer Science}, pages 198--212. Springer, 2002.

\bibitem{grumbach}
S.~Grumbach, P.~Rigaux, and L.~Segoufin.
\newblock Spatio-temporal data handling with constraints.
\newblock In R.~Laurini, K.~Makki, and N.~Pissinou, editors, {\em Proceedings
  of the 6th International Symposium on Advances in Geographic Information
  Systems (ACM-GIS'98)}, pages 106--111, 1998.

\bibitem{ge-00}
R.~H. G{\"u}ting, M.~H. Bohlen, M.~Erwig, C.~S. Jensen, N.~A. Lorentzos,
  M.~Schneider, and M.~Vazirgiannis.
\newblock A foundation for representing and querying moving objects.
\newblock {\em ACM Transactions on Databases Systems}, 25:1--42, 2000.

\bibitem{gvv-jcss}
M.~Gyssens, J.~Van den Bussche, and D.~Van Gucht.
\newblock Complete geometric query languages.
\newblock {\em Journal of Computer and System Sciences}, 58(3):483--511, 1999.

\bibitem{kh-gis04}
S.~Haesevoets and B.~Kuijpers.
\newblock Time-dependent affine triangulation of spatio-temporal data.
\newblock In {\em 12th ACM International Workshop on Geographic Information
  Systems, ACM-GIS}, pages 57--66, 2004.

\bibitem{sym-diff}
M.~Hagedoorn and R.~C. Veldkamp.
\newblock Reliable and efficient pattern matching using an affine invariant
  metric.
\newblock {\em International Journal of Computer Vision}, 31:203--225, 1999.

\bibitem{pattern-hausdorff}
D.P. Huttenlocher, G.A. Klauderman, and W.J. Rucklidge.
\newblock Comparing images using the {H}ausdorff distance.
\newblock {\em IEEE Transactions on Pattern Analysis and Machine Intelligence},
  15:850--863, 1998.

\bibitem{hk-00}
B.~Kuijpers and S.~Haesevoets.
\newblock Closure properties of classes of spatio-temporal objects under
  boolean set operations.
\newblock  \cite{time00}, pages 79--86.

\bibitem{kpv-00}
B.~Kuijpers, J.~Paredaens, and D.~Van Gucht.
\newblock Towards a theory of movie database queries.
\newblock In {\em Proceedings of the 7th International Workshop on Temporal
  Representation and Reasoning\/} \cite{time00}, pages 95--102.

\bibitem{nielson}
G.~Nielson.
\newblock A characterization of an affine invariant triangulation.
\newblock In G.~Farin, H.~Hagen, and H.~Noltemeier, editors, {\em Geometric
  Modelling, Computing Supplementum 8}, pages 191--210, 1993.

\bibitem{rootfinding}
E.~Novak and K.~Ritter.
\newblock Some complexity results for zero finding for univariate functions.
\newblock {\em Journal of Complexity}, 9:15--40, 1993.

\bibitem{pods-94}
J.~Paredaens, J.~Van den Bussche, and D.~Van Gucht.
\newblock Towards a theory of spatial database queries.
\newblock In {\em Proceedings of the 13th ACM Symposium on Principles of
  Database Systems}, pages 279--288. ACM Press, 1994.

\bibitem{cdbook}
J.~Paredaens, G.~Kuper, and L.~Libkin, editors.
\newblock {\em Constraint databases}.
\newblock Springer-Verlag, 2000.

\bibitem{req}
D.~Pfoser and N.~Tryfona.
\newblock Requirements, definitions and notations for spatiotemporal
  application environments.
\newblock In R.~Laurini, K.~Makki, and N.~Pissinou, editors, {\em Proceedings
  of the 6th International Symposium on Advances in Geographic Information
  Systems (ACM-GIS'98)}, pages 124--130, 1998.

\bibitem{reveszbook}
P.~Revesz.
\newblock {\em Introduction to Constraint Databases}.
\newblock Springer-Verlag, 2002.

\bibitem{weakpersp}
L.G. Roberts.
\newblock Machine perception of three-dimensional solids.
\newblock {\em J.T. Tippet, editor, Optical and Electro-optical Information
  Processing}, 1965.

\bibitem{geom-hashing}
J.T.~Schwartz Y.~Lamdan and H.J. Wolfson.
\newblock Affine-invariant model-based object recognition.
\newblock {\em IEEE Journal of Robotics and Automation}, 6:578--589, 1990.

\end{thebibliography}
